\documentclass[preprints,article,accept,pdftex,oneauthor]{Definitions/mdpi}

\firstpage{1}
\pubvolume{1}
\issuenum{1}
\articlenumber{0}
\pubyear{2024}
\copyrightyear{2024}
\datereceived{ }
\daterevised{ } 
\dateaccepted{ }
\datepublished{ }
\hreflink{https://doi.org/} 

\usepackage{rotating}

\usepackage{color, colortbl}
\definecolor{Gray}{gray}{0.9}
\definecolor{LightCyan}{rgb}{0.88,1,1}

\Title{Long-term usage of the off-grid photovoltaic system with lithium-ion battery-based energy storage system on high mountains: A case study in Payiun Lodge on Mt. Jade in Taiwan}

\TitleCitation{Title}

\Author{Hsien-Ching Chung $^{1}$\orcidA{}}

\AuthorNames{Hsien-Ching Chung}

\AuthorCitation{Chung, H.C.}

\address{%
$^{1}$ \quad Department of Research and Design, Super Double Power Technology Co., Ltd., Changhua City, Changhua County 500042, Taiwan; hsienching.chung@gmail.com
}

\corres{Correspondence: hsienching.chung@gmail.com (H.C. Chung)}

\abstract{
Energy supply on high mountains remains an open issue since grid connection is unavailable.
In the past, diesel generators with lead-acid battery energy storage systems (ESSs) are applied in most cases.
Recently, photovoltaic (PV) system with lithium-ion (Li-ion) battery ESS is an appropriate method for solving this problem in a greener way.
In 2016, an off-grid PV system with Li-ion battery ESS has been installed in Paiyun Lodge on Mt. Jade (the highest lodge in Taiwan).
After operation for more than 7 years, the aging problem of the whole electric power system becomes a critical issue for long-term usage.
In this work, a method is established for analyzing the massive energy data (over 7 million rows) and estimating the health of the Li-ion battery system, such as daily operation patterns as well as C-rate, temperature, and accumulated energy distributions.
The accomplished electric power improvement project dealing with the power system aging is reported.
Based on the long-term usage experience, a simple cost analysis model between lead-acid and Li-ion battery systems is built, explaining that the expensive Li-ion batteries can compete with the cheap lead-acid batteries for long-term usage on high mountains.
This case study provides engineers and researchers a fundamental understanding of the long-term usage of off-grid PV ESSs and engineering on high mountains.
}

\keyword{Lithium-ion battery; power improvement; high mountain; cost analysis; Paiyun Lodge}

\begin{document}

\section{Introduction}

Lithium-ion (Li-ion) batteries are one of the most widely used rechargeable batteries in the world.
Compared to other traditional secondary batteries, e.g., lead-acid batteries, nickel-cadmium, and nickel-metal hydride, Li-ion batteries exhibit higher operating voltages, higher energy density, low self-discharging, and lower maintenance requirements~\cite{J.PowerSources195(2010)2419B.Scrosati, Nature451(2008)652M.Armand, Nature414(2001)359J.M.Tarascon}.
Recently, International Energy Agency (IEA) reported that electric vehicle (EV) markets show exponential growth as sales exceeded 10 million in 2022. A total of 14\% of all new EVs sold were electric in 2022, up from around 9\% in 2021 and less than 5\% in 2020.
EV Li-ion battery demand increased by about 65\% to 550 GWh in 2022, from about 330 GWh in 2021~\cite{GlobalEVOutlook2023IEA}.
The advance of renewable energy supply and EV industries enhance the application of Li-ion batteries from small-scale 3C (computing, communication, and consumer) products~\cite{Adv.Mater.Technol.8(2023)2200459Z.Bassyouni, Sci.Adv.12(2023)eadg5135J.J.Lodico} to large-scale battery energy storage systems (BESSs)~\cite{J.EnergyStorage87(2024)111508N.Collath, ProcessSaf.Prog.43(2024)357V.Goldsmith,
Energies17(2024)1019G.L.Trombetta} and high-power mobile energy sources.



The installation of the grid-scale Li-ion battery (100 MW, 129 MWh from Tesla and Neoen) in South Australia in 2017 has demonstrated the capability of batteries in electric grid stabilization~\cite{Energy173(2019)647F.Keck, Water-EnergyNexus1(2018)66J.C.Radcliffe}.
After several weeks, when the Loy Yang coal-fired power plant in Victoria failed, leading to a power shortage, the backup Li-ion battery connected in and sent about 100 MW into the grid within 140 ms~\cite{Renew.Sustain.EnergyRev.111(2019)145Y.Ye, CICED(2018)2895H.Zeng}, responding even more quickly than the coal-fired backups that were supposed to provide emergency power.
That shock emitter- and absorber-type capacities help us to avoid a blackout that would otherwise occur.
The batteries can provide inertia services and rapid frequency responses (e.g., frequency control ancillary services; FCAS) to the grid, paving the way for potential regulatory modifications and revenue streams to incentivize further grid-scale energy storage systems (ESSs)~\cite{Renew.Sustain.EnergyRev.120(2020)109662D.Fernandez-Munoza, EnergyTechnol.7(2019)1900791M.Pagliaro, J.Mod.PowerSyst.CleanEnergy6(2018)1141A.Aziz}.
In its first year, the grid-scale Li-ion battery facility saved nearly 40 million USD as well as helped to balance and stabilize the region's unstable grid.
Large-scale battery systems showed that they will play an important role in supplying electric networks.



Energy supply on high mountains remains an open issue for solving.
Grid connection is impractical, since the establishing and maintenance costs are too high.
On the other hand, there are so many environmental problems, such as destroying the natural ecology on the mountain.
Diesel generators have been the energy supplier on high mountains for a long time despite air pollution and noise problems.
Owing to the development of renewable energy (such as solar, wind, and water), the usage of diesel generators is reduced, lowering the emission of greenhouse gas (GHG)~\cite{Renew.Sustain.EnergyRev.197(2024)114412C.K.Saha, Sustain.EnergyGridsNetw.38(2024)101330J.Yumbla, J.Mar.Sci.Eng.12(2024)843B.B.Basic}.
However, most renewable energy possesses intermittent features due to their fluctuating nature.
To deal with the intermittent problem of renewable energy, ESSs are necessary~\cite{IEEETrans.SmartGrid6(2015)124K.Rahbar, IEEETrans.Sustain.Energy1(2010)117S.Teleke}.
In the past, lead-acid batteries were heavily used as ESSs, accompanied by many environmental issues, e.g., poisoning, leaks, contaminating the environment, and damaging the ecosystem~\cite{Ecol.Indic.47(2014)210G.N.Liu, J.Hazard.Mater.250-251(2013)387X.F.Zhu}.
Recently, Li-ion batteries as a greener alternative start to replace lead-acid batteries in the ESSs.





\begin{figure}
\centering
\includegraphics[width=\linewidth,keepaspectratio]{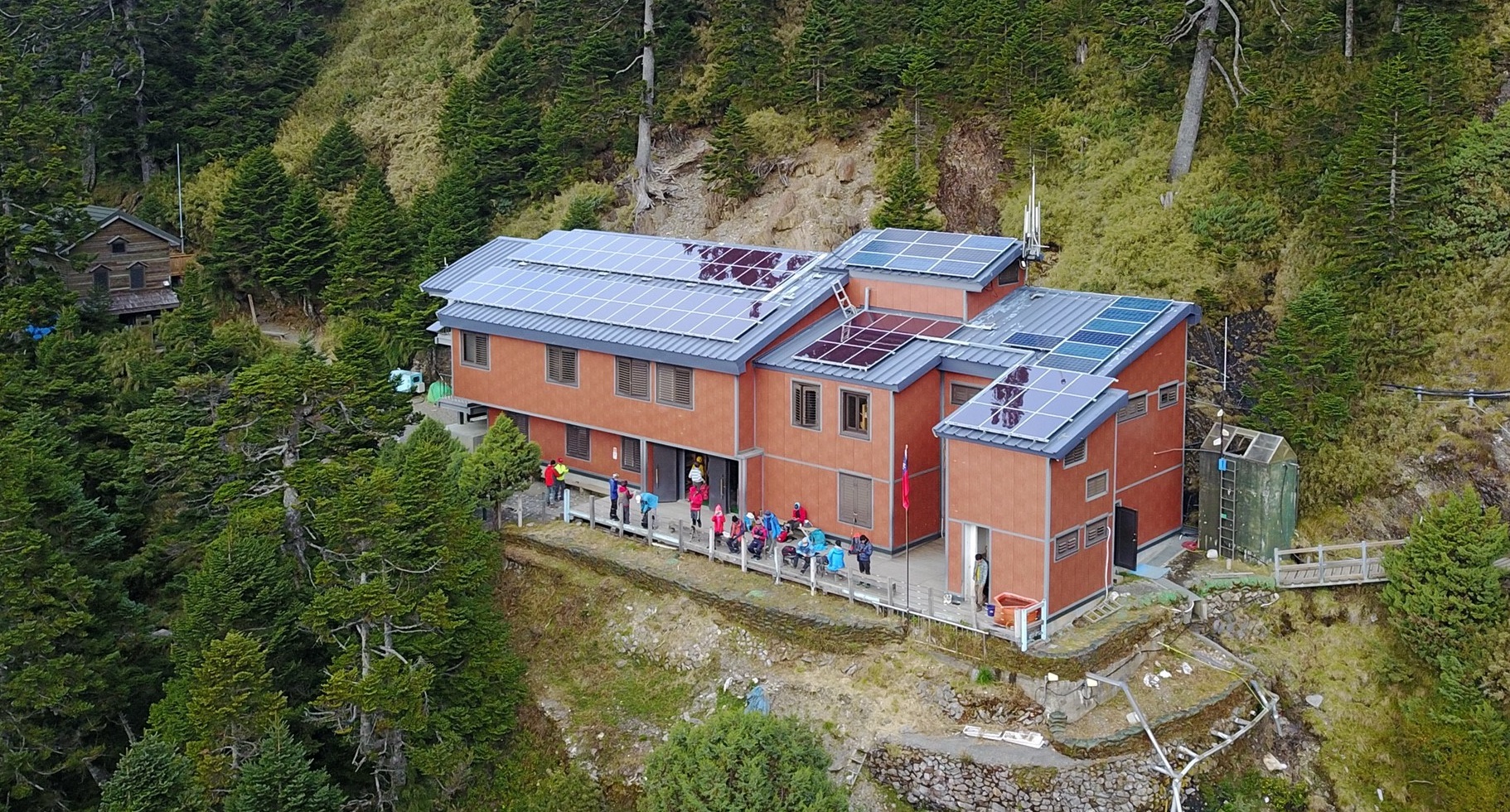}
\caption{
Paiyun Lodge, the highest lodge in Taiwan, is at an altitude of 3402 m (11161 ft), located 2.4 km (1.5 mi) below the west slope of the main peak of Mt. Jade.
Photographer: Yu-Chun Lin.
}
\label{fig:PaiyunLodge}
\end{figure}

Since 2016, the off-grid photovoltaic (PV) ESS has been installed in Paiyun Lodge, the highest mountain lodge in Taiwan (as shown in Fig.~\ref{fig:PaiyunLodge}).
In the system, solar panels provide intermittent energy generation and the Li-ion battery ESS serves as an energy reservoir.
In 2020, a system aging estimation project on the long-term usage of the solar panels and ESS were initialized, reporting that the whole system is aging but most of the system remains healthy for long-term usage.
Some repair and optimization suggestions were given in 2021.
In 2022, we have accomplished the electric power improvement project, dealing with the system aging problems.

\begin{figure}
\centering
\includegraphics[width=\linewidth,keepaspectratio]{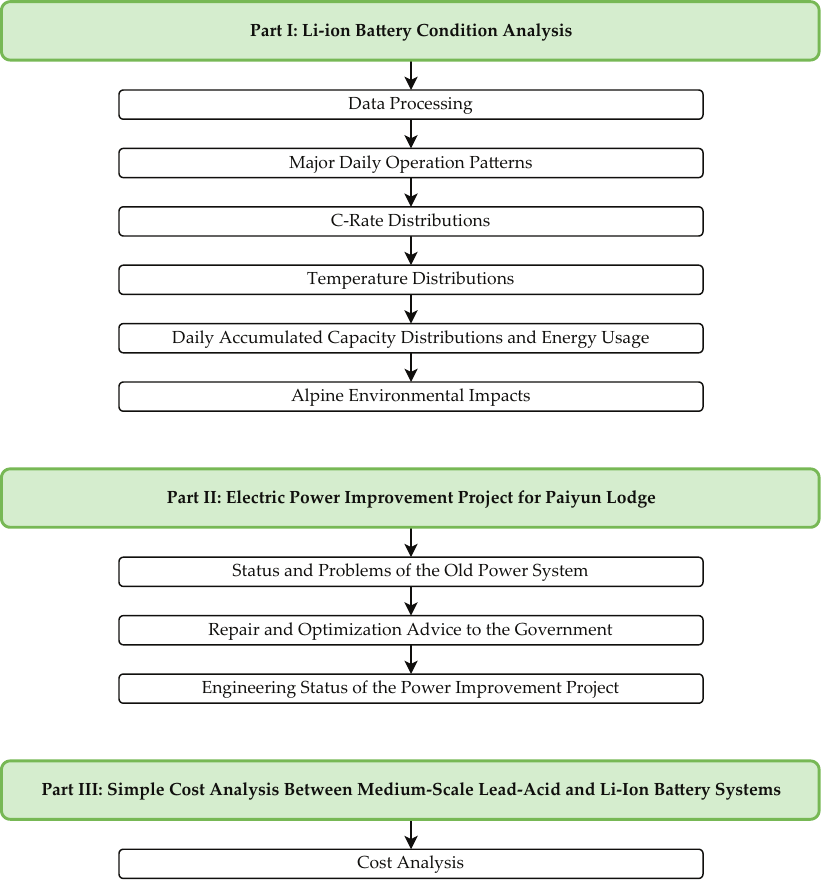}
\caption{
Methodology scheme of the work.
Three parts are contained, i.e., part I: Li-ion battery condition analysis, part II: electric power improvement project for Paiyun Lodge, and part III: simple cost analysis between medium-scale lead-acid and Li-ion battery systems.
}
\label{fig:MethodologyScheme}
\end{figure}

The research methodology scheme of this work including three parts is shown in Fig.~\ref{fig:MethodologyScheme}.
In the first part (Sec.~\ref{sec:BatteryConditionAnalysis}), the energy data of the Li-ion battery system for four years is analyzed.
Analysis of such massive data is not an easy task (more than 7 million rows).
Especially, the data is recorded under the actual operation instead of regular data from cycle life tests in the laboratory.
A method is established for analyzing the energy data during operation and estimating the health of the Li-ion battery system, including daily operation patterns as well as C-rate, temperature, and accumulated energy distributions.
In the second part (Sec.~\ref{sec:ElectricPowerImprovementProject}), the status and problems of the old electric power system of Paiyun Lodge are described.
The engineering status of the electric power system is reported, such as (1) reform and optimization of existing Li-ion battery cabinets, (2) PV inverter system reformation and optimization, (3) reorganization of distribution boxes and power line adjustment, (4) a self-developed cloud energy management system (EMS) is installed to remotely monitor the off-grid PV ESS.
In the third part (Sec.~\ref{sec:BriefCostAnalysis}), a simple cost model is built for comparing the cost between lead-acid and Li-ion battery systems.
This model reasonably explains that the expensive Li-ion batteries can compete with the cheap lead-acid batteries for long-term usage on high mountains.
Above all, this study gives engineers and researchers a fundamental understanding of a real case of long-term usage of off-grid PV ESSs on high mountains.









\section{Li-ion Battery Condition Analysis}
\label{sec:BatteryConditionAnalysis}



The Li-ion battery system with a nominal voltage of 48 V and capacity of 250 Ah is made by a 16S5P configuration, i.e., 5 batteries connected in parallel as a group and 16 groups of batteries connected in series as a whole batter system.
For long durability and thermal stability, lithium iron phosphate (LFP) batteries are chosen as battery cells in the system~\cite{Sci.Data8(2021)165H.C.Chung, BookChChung2021UL1974, BookLinFirstPrinciplesCathodeElectrolyteAnodeBatteryMaterials, MJTE860(2020)35H.C.Chung, J.TaiwanEnergy6(2019)425H.C.Chung, FMEAofLFPBatteryModule2018Chung}.
The data was obtained from the battery management system (BMS)~\cite{J.EnergyStorage88(2024)111567J.Patel, J.Radiat.Res.Appl.Sci.17(2024)100927M.Yagci, J.EnergyStorage86(2024)111327M.Tekin}, including a total voltage ($V_{tot}$), 16 cell voltages ($V_n, n=1 \sim 16$), a current ($I$), and 4 temperatures ($T_n, n=1 \sim 4$).
The data was recorded from Oct. 14th, 2016 to Jul. 20th, 2020 (about 1376 days, across 5 years).
There are over 7 million rows of data with data time intervals of about 15 seconds.
In the following subsections, the data processing, major daily operation pattern, C-rate distribution, temperature distribution, capacity distribution, and energy usage of the battery system are discussed.

\subsection{Data Processing}

Data processing is crucial before data analysis. The data obtained from BMS is stored in a CSV (comma-separated values) file every day. There are about 1400 CSV files in the data storage. The time duration between data points is about 15 seconds. The accuracy of data and time plays an important role in the data processing. If the time is obtained from the standalone BMS without network calibration, the timeline of the data should be checked first. In this case, the time was obtained based on the passive device (i.e., a crystal Oscillator), and there was no network connected for time calibration once a day or once a week. After drawing the first and last CSV file, a huge time shift of about 15000 seconds was found.

\begin{figure}
\begin{adjustwidth}{-\extralength}{0cm}
\centering
\includegraphics[width=\linewidth,keepaspectratio]{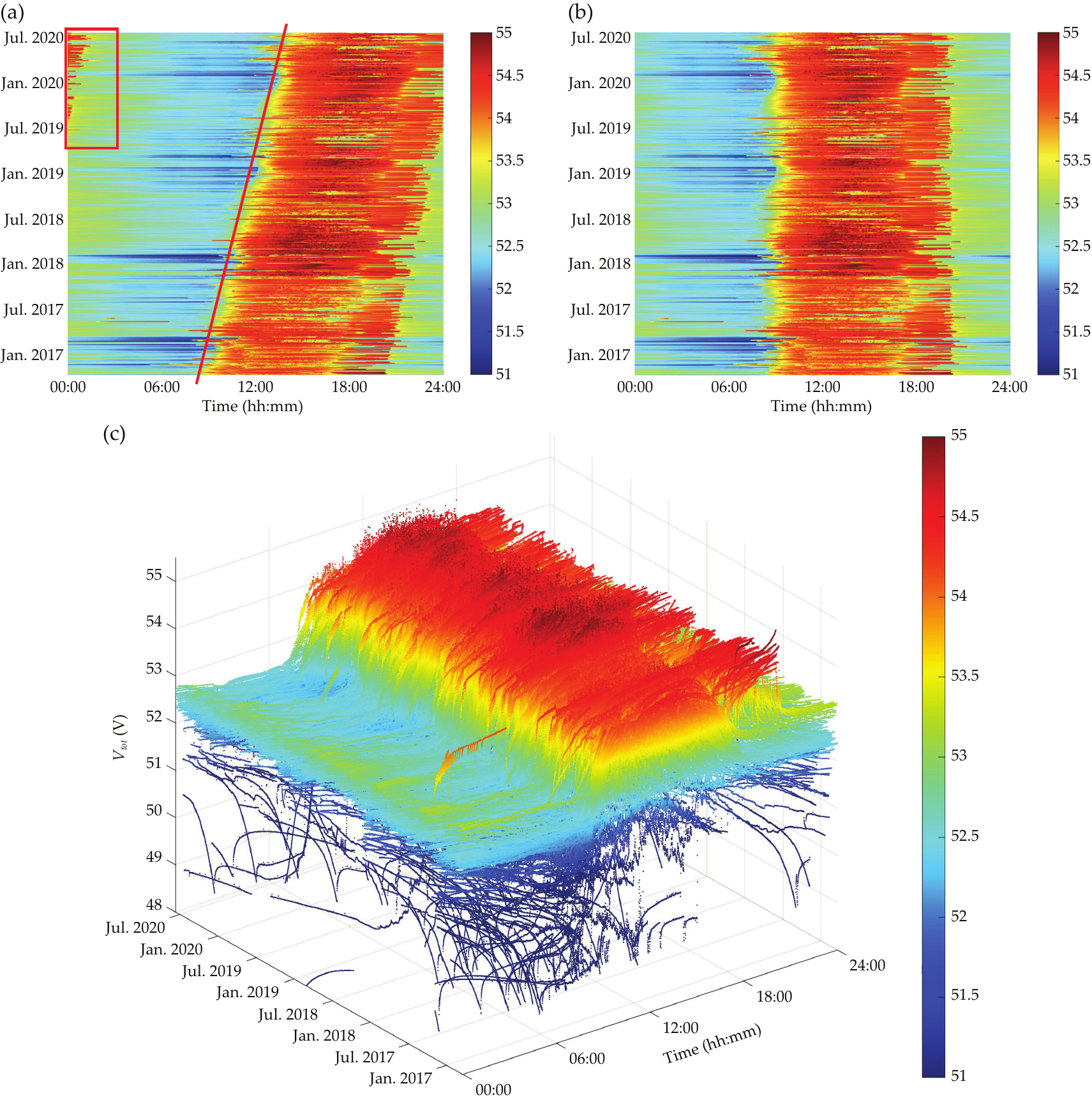}
\end{adjustwidth}
\caption{
Data processing.
(a) Colormap of total voltage obtained from the source data.
The oblique red line indicates a time shift (about 15000 seconds) in the source data, causing a serious data shift.
It's obvious that the data in the red rectangle should be on the right side of the figure.
(b) Colormap of total voltage obtained from the time-adjusted data.
After time adjustment, the data exhibits a clear daily pattern.
(c) 3D plot of total voltage obtained from the time-adjusted data.
}
\label{fig:DataProcessing}
\end{figure}

It's difficult to deal with 1400 CSV files one by one manually, e.g., using Microsoft$^\circledR$ Excel or some lightweight scientific drawing software. In this work, MathWorks$^\circledR$ MATLAB was used. The first step was to combine the data in 1400 CSV files, i.e., putting all datasets in a single database. Then, drawing a daily colormap to see the time shift condition. Because of the daily routine operation of the ESS, the data is predicted to exhibit a roughly daily pattern on a large time scale. The time shift issue can be solved by adjusting the time until the data exhibits a roughly daily pattern. As shown in Fig.~\ref{fig:DataProcessing}(a), the time shift caused an obvious pattern shift of total voltage. In Jan. 2017, the red region began at about 8 o’clock, while in Jul. 2020, the red region began at about 12 o’clock. The time shift was a continuous process, exhibiting a parallelogram of red region. The red line on the left side of the parallelogram can serve as an indicator of time shift. The more sloping the red line is, the more serious the time shift is. Some data on the right side were shifted to the left side (indicated by the red rectangle). The time shift issue can be solved by linear adjustment as shown in Fig.~\ref{fig:DataProcessing}(b). The red region becomes a rectangle, indicating a roughly daily pattern. A three-dimensional (3D) plot figure for showing the variation of total voltage is given in Fig.~\ref{fig:DataProcessing}(c).

Other technical suggestions for efficient large-scale data processing are given. (1) Using parallel computing technique. In this study, 12 cores in the CPU were used simultaneously for data calculations. (2) Use programmable drawing software. There are about 1400 figures in this work. Writing a program to perform batch drawing, instead of drawing figures one by one manually.

\subsection{Major Daily Operation Patterns of the Battery System}

It is difficult to look at such a huge amount of data (over 7 million rows) in detail.
Especially, the data obtained from real cases are not similar to the data from professional instruments.
There is no explicit constant current (CC) discharge mode and constant current-constant voltage (CC-CV) charge mode for analysis~\cite{Sci.Data8(2021)165H.C.Chung,J.PowerSources262(2014)129T.Waldmann}.
We need some analyzing skills.
Looking at the data in units of days, we found that there are several major daily operation patterns.
These major patterns are highly related to the amount of solar power generated and the operating conditions of the diesel generators as shown in Table~\ref{tab:ESS_Patterns}.
There are three solar power conditions (high, medium, and low) and two generator conditions (ON/OFF).
For example, the major pattern 1 is in high solar power and generator OFF conditions.

\begin{table}
\centering
\caption{Conditions for major daily operation patterns of the battery system.
Four major daily operation patterns are listed based on various conditions of solar power and generator.
}
\begin{tabular}{c | c c}
\hline
 & Generator OFF & Generator ON \\
\hline
High solar power   & Pattern 1          & No major pattern  \\
Medium solar power & Pattern 2          & Pattern 3         \\
Low solar power    & No major pattern   & Pattern 4         \\
\hline
\end{tabular}
\label{tab:ESS_Patterns}
\end{table}

\begin{figure}
\begin{adjustwidth}{-\extralength}{0cm}
\centering
\includegraphics[width=\linewidth,keepaspectratio]{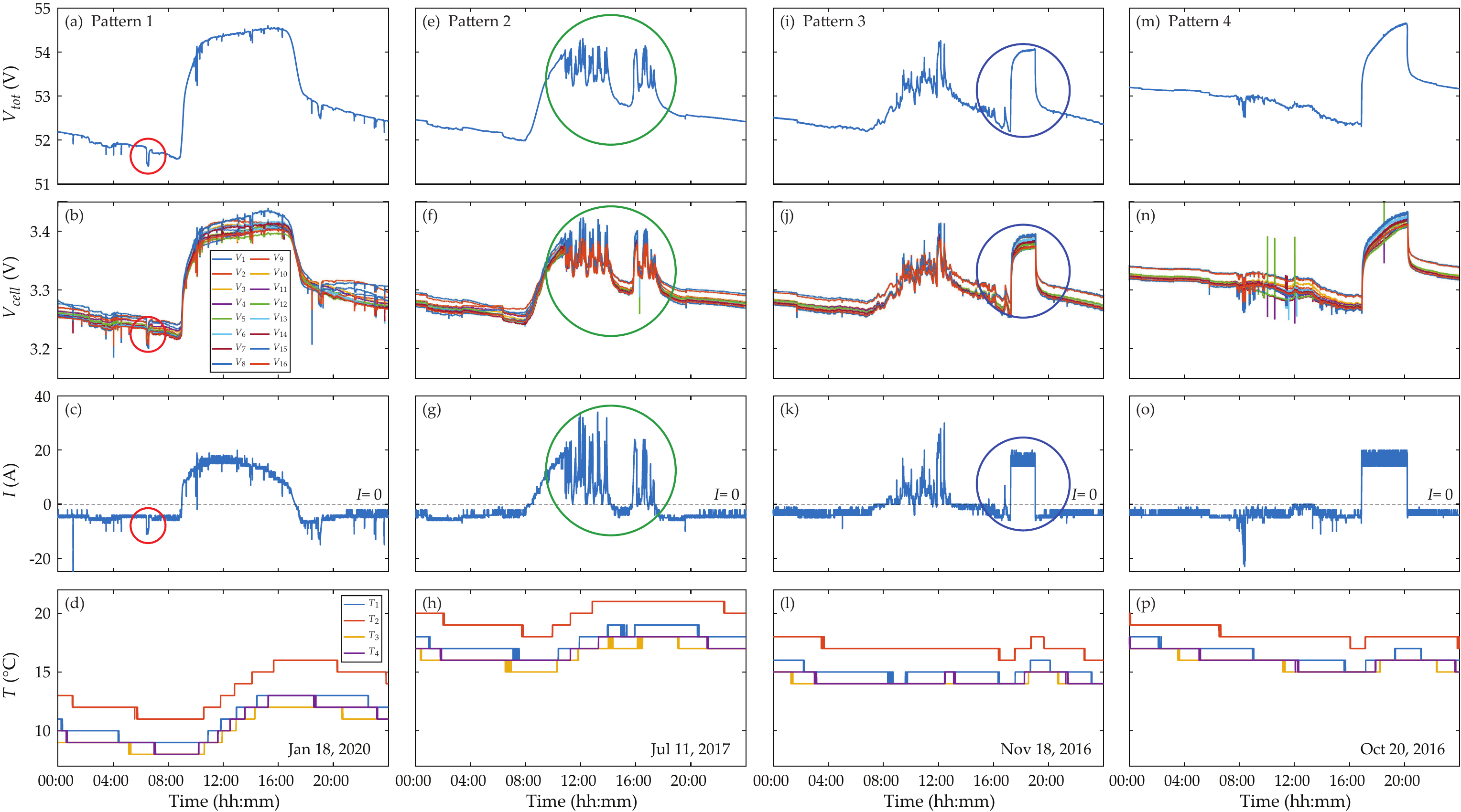}
\end{adjustwidth}
\caption{
Four major daily operation patterns of the battery system (corresponding pattern conditions are listed in Table~\ref{tab:ESS_Patterns}).
Patterns 1, 2, 3, and 4 are shown in (a)--(d), (e)--(h), (i)--(l), (m)--(p), respectively.
Each pattern contains the total voltage ($V_{tot}$), 16 cell voltages ($V_n, n=1 \sim 16$), current ($I$), and 4 temperatures ($T_n, n=1 \sim 4$).
The level of zero current ($I = 0$) is denoted by dashed lines, and the charge (discharge) current is a positive (negative) value.
The red circles highlight the instant voltage drop and high current of the battery system.
The green circles highlight many instant voltage and current drops under the charge state of the battery system.
The blue circles mark the characteristic voltage and current variations caused by the diesel generator.
The date is displayed in the lower right corner of each pattern.
}
\label{fig:ESS_Major_Patterns}
\end{figure}

The massive data exhibits daily patterns.
Four major daily patterns are observed during system operation (while not all conditions exhibit explicit patterns).
The selected data as example cases are illustrated in Fig.~\ref{fig:ESS_Major_Patterns}.
For pattern 1 (high solar power and generator OFF conditions), the total voltage ($V_{tot}$) of the battery system exhibits a smooth curve with a bump in the daytime (Fig.~\ref{fig:ESS_Major_Patterns}(a)).
During the nighttime (about 17:00--09:00), no solar power is generated.
The battery system is in the discharge state with current ($I$) about 5 A (Fig.~\ref{fig:ESS_Major_Patterns}(c)).
$V_{tot}$ and $V_{cell}$ gradually decrease (Fig.~\ref{fig:ESS_Major_Patterns}(a) and (b)).
During the daytime (about 09:00--17:00), the generated solar power is sufficient for the load and the battery system.
The battery system is in the charge state with $I$ of a maximum of $\sim$20 A (Fig.~\ref{fig:ESS_Major_Patterns}(c)).
$V_{tot}$ and $V_{cell}$ gradually increase to values of 54.5 V and $\sim$3.4 V, respectively (Fig.~\ref{fig:ESS_Major_Patterns}(a) and (b)).
$V_{tot}$ varies under the safety operation range of about 48--56 V and $V_{cell}$s vary under the safe operation range of LFP battery cells about 3--3.5 V~\cite{Sci.Data8(2021)165H.C.Chung, BookChChung2021UL1974}.
Some small voltage drops in $V_{tot}$ and $V_{cell}$ are due to instantaneous high power consumptions of the load (red circles in Fig.~\ref{fig:ESS_Major_Patterns}(a) and (b)), which can be verified by the large $I$ at the corresponding time (red circle in Fig.~\ref{fig:ESS_Major_Patterns}(c)).

For pattern 2 (medium solar power and generator OFF conditions) as shown in Figs.~\ref{fig:ESS_Major_Patterns}(e)--(h), the voltage and current behaviors during the nighttime are in the discharge state and the same as those of pattern 1.
During the daytime, the generated solar power is sufficient for the load, but not sufficient for the battery system.
The battery system is still in the charge state, while many instant current and voltage drops are exhibited (green circles in Figs.~\ref{fig:ESS_Major_Patterns}(e)--(g)), reflecting the intermittent power generation characteristics of renewable energy.
For pattern 3 (medium solar power and generator ON conditions) as shown in Figs.~\ref{fig:ESS_Major_Patterns}(i)--(l), the voltage and current behaviors are similar to those of pattern 2.
During the daytime, insufficient solar power cannot fulfill the battery system, many instant current and voltage drops are exhibited in the charge state (Figs.~\ref{fig:ESS_Major_Patterns}(i)--(k)).
To solve the energy shortage, the backup generator is turned on with a constant charge current of about 18 A (blue circle in Fig.~\ref{fig:ESS_Major_Patterns}(k)), and the voltages gradually are charged to saturated values (blue circles in Figs.~\ref{fig:ESS_Major_Patterns}(i) and (j)).

For pattern 4 (low solar power and generator ON conditions) as shown in Figs.~\ref{fig:ESS_Major_Patterns}(m)--(p), the voltage and current exhibit a gradual decrease with a large bump.
During the daytime, low or no solar power is generated, causing the battery system in the discharge state, i.e., the voltages gradually decrease (Figs.~\ref{fig:ESS_Major_Patterns}(m) and (n)) and the current is around 5 A (Fig.~\ref{fig:ESS_Major_Patterns}(o)).
Later, the backup generator is turned on to charge the battery system.
The battery system operation patterns give us a convenient way to recognize several normal operation modes of the battery system and quickly figure out abnormal conditions.

\subsection{C-Rate Distributions of the Battery System}

C-rate ($C_R$) is one of the key and simple factors to analyze the charge and discharge conditions of the battery system, where $C_R \equiv I/Cap_N$ is the current $I$ per unit of nominal ampere-hour capacity $Cap_N$~\cite{Electrochem.Soc.Interface27(2018)42J.Jacob,Sci.Data8(2021)165H.C.Chung}.
The current is usually expressed in terms of C-rate, i.e., the current is normalized to the nominal capacity of the battery.
The increase of charge and discharge C-rate causes ohmic heating with a temperature rising, resulting in Li-ion batteries to degrade faster.
Chen et al. designed a calorimeter to measure the thermal behaviors of LFP prismatic batteries at $C_R = 0.25$--3 h$^{-1}$ and found that the heat generation rate increases with C-rate~\cite{J.PowerSources261(2014)28K.Chen}.
Drake et al. studied the heat generation rate of a 26650 LFP battery by heat flux measurements under the discharge C-rate ranging from $C_R = 1$--9.6 h$^{-1}$.
At a high C-rate of $C_R = 9.6$ h$^{-1}$, the surface temperature rise climbed to 40$^\circ$C and even reached the temperature safety limit before finishing the entire discharge process~\cite{J.PowerSources285(2015)266S.J.Drake}.
Chung et al. found that high C-rate charging ($C_R > 5$ h$^{-1}$) is unsuitable for batteries and might lead to thermal runaway and fire hazards~\cite{J.Therm.Anal.Calorim.127(2017)809Y.H.Chung}.
Wu et al. investigated the relationship among the temperature, discharge C-rate, and cycle life of pouch cells.
At room temperature, the larger the C-rate, the faster the capacity fades and the larger the ohmic resistance~\cite{J.Electrochem.Soc.164(2017)A1438Y.Wu}.
The solid electrolyte interface (SEI) layer grows with temperatures over 25$^\circ$C, leading to faster irreversible capacity fade~\cite{J.PowerSources262(2014)129T.Waldmann, Sci.Rep.5(2015)12967F.Leng}.
Besides the capacity fades, a high C-rate can also accelerate the aging process of cell components in a Li-ion battery and shorten the cycle life~\cite{Int.J.Electr.PowerEnergySyst.107(2019)438S.Saxena}.
For example, the chemical composition of the surface of the current collector changes significantly~\cite{J.PowerSources335(2016)189L.Somerville}, the decomposition of electrolytes (e.g., LiPF$_6$) is accelerated~\cite{Prog.Nat.Sci.28(2018)653S.Ma}, and massive ethylene (C$_2$H$_4$) evolution occurs~\cite{J.PowerSources477(2020)228968U.Mattinen}.







\begin{figure}
\centering
\includegraphics[width=0.8\linewidth,keepaspectratio]{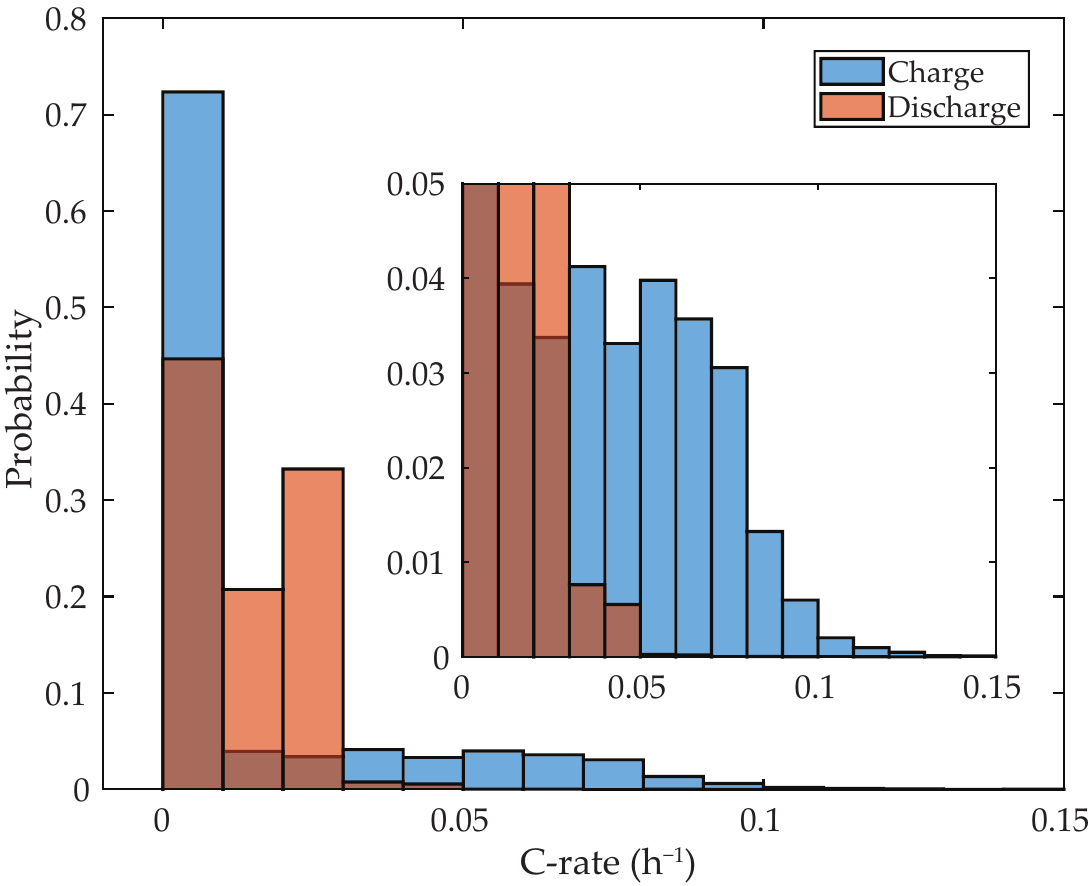}
\caption{
Time distribution of C-rate during battery system operation.
The probability of charge and discharge C-rate are marked by blue and red bars, respectively.
Inset shows the low-probability region.
The battery system operated under a C-rate below $C_R < 0.15$ h$^{-1}$, which lower than the suggested value of $C_R < 0.25$ h$^{-1}$, advancing system stability and prolonging the life of the system.
Time range of data: 2016-10-14$\sim$2020-07-20. (Total: 1376 days.)
}
\label{fig:C_rate_histogram}
\end{figure}

Unlike power batteries used in EVs, which have instantaneous high power requirements ($C_R > 1$ h$^{-1}$), most battery systems in the stationary application are designed to normally operate at a low C-rate ($C_R < 0.25$ h$^{-1}$) with an instantaneous high-power output mode at $C_R < 1$ h$^{-1}$ for safety and service life extension~\cite{J.Electrochem.Soc.169(2022)030509M.Bozorgchenani,BookChChung2021EngIntePotentialAppOutlooksLiIonBatteryIndustry,engrxiv2020OutlooksLiIonBatteriesH.C.Chung}.
For example, the grid-scale Li-ion battery (100 MW, 129 MWh from Tesla and Neoen) has been
installed in Hornsdale, South Australia in 2017~\cite{Energy173(2019)647F.Keck,Water-EnergyNexus1(2018)66J.C.Radcliffe}.
As backup energy storage to the grid, when the Loy Yang coal-fired power plant in Victoria failed, leading to a power shortage, a power of 100 MW was input into the national electricity grid within 140 ms~\cite{Renew.Sustain.EnergyRev.111(2019)145Y.Ye,CICED(2018)2895H.Zeng}.
In this case, the instantaneous output C-rate was about 0.78 h$^{-1}$.
For the discussed battery system, the time distribution of C-rate during battery system operation is shown in Fig.~\ref{fig:C_rate_histogram}.
Almost all values (over 99.9\%) of C-rate during charge are under 0.1 h$^{-1}$, and more than 70\% of the time is at $C_R < 0.01$.
On the other hand, the C-rate during discharge almost distributes under 0.05 h$^{-1}$, and over 95\% of the time is at $C_R < 0.03$.
The battery system operation is under a low C-rate condition with the values of C-rate lower than the normal design C-rate, i.e., $C_R < 0.25$ h$^{-1}$.
The low C-rate condition can avoid excessive voltage differences among battery cells and heat concentration on the current collectors, ensuring system stability and prolonging the life of the system.


\subsection{Temperature Distributions of the Battery System}

\begin{figure}
\centering
\includegraphics[width=0.8\linewidth,keepaspectratio]{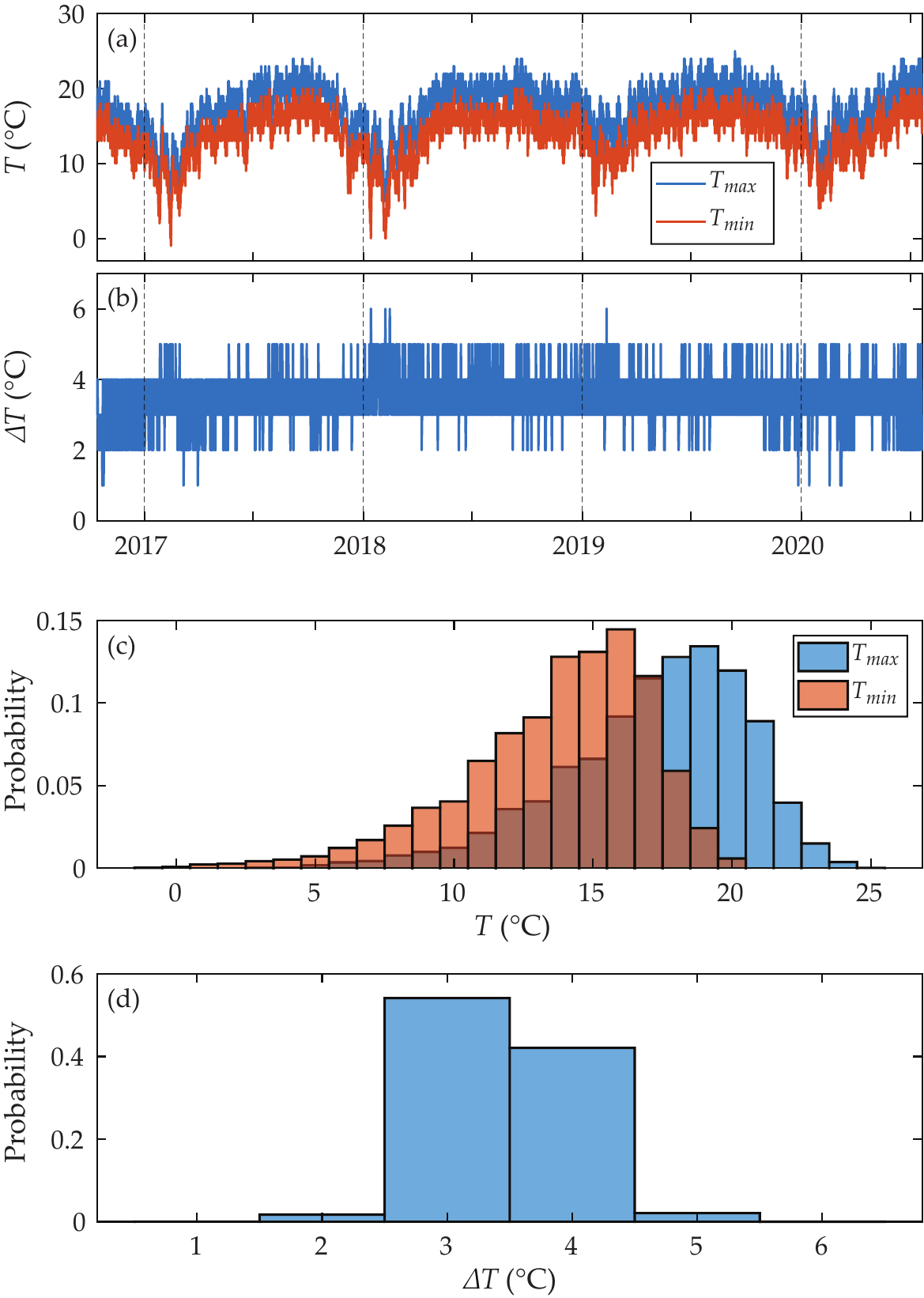}
\caption{
Temperature distribution of the battery system.
(a) The variation of maximum and minimum temperatures over time, where $T_{max}$ and $T_{min}$ are indicated by blue and red curves, respectively.
(b) The variation of temperature difference ($\Delta T$) over time.
(c) The distribution of maximum and minimum temperatures.
(d) The distribution of temperature difference.
Time range of data: 2016-10-14$\sim$2020-07-20. (Total: 1376 days.)
}
\label{fig:Temperature}
\end{figure}

Operation temperature and temperature distribution have a large impact on the performance, lifespan, and safety of the battery system.
The acceptable operating temperature for Li-ion batteries ranges from -20 to 60$^\circ$C~\cite{J.Chem.Thermodyn.46(2012)80A.Vayrynen}.
To maintain better performance, a narrow temperature range of about 15--35$^\circ$C is recommended ~\cite{EnergyConvers.Manag.150(2017)304H.Liu, Appl.Therm.Eng.94(2016)846D.Chen}.
Temperatures outside the desired operation range will lead to fast capacity fade.
Moreover, when the cell temperature exceeds the threshold of the thermal runaway point, chain electrochemical reactions might yield battery cells to burn and explode~\cite{NanoEnergy27(2016)313J.Sun}.
The operation temperature of the discussed battery system ranging from 0 to 25$^\circ$C is shown in Fig.~\ref{fig:Temperature}(a).
The temperature variation exhibits an annual pattern, reflecting seasonal climate characteristics, i.e., high temperature in summer and low temperature in winter.
More than half of the time is in the recommended range of operation temperature (Fig.~\ref{fig:Temperature}(c)).

The temperature difference ($\Delta T=T_{max}-T_{min}$) among the battery cells in a pack should be as low as possible for better performance of the battery system, where $T_{max}$ and $T_{min}$ are the maximum and minimum temperatures of the battery cells in pack, respectively.
Such condition is called temperature uniformity (or thermal balance) and the desired value to be less than 5$^\circ$C~\cite{HeatTrans.51(2022)7540S.Gungor, J.PowerSources436(2019)226879Z.Liao, J.PowerSources275(2015)742C.Lin, J.PowerSources196(2011)5685R.Mahamud, J.PowerSources110(2002)377A.A.Pesaran}.
For the discussed battery system, the $\Delta T$ of all time is almost under 5$^\circ$C and up to 6$^\circ$C several times (Fig.~\ref{fig:Temperature}(b)).
As illustrated in Fig.~\ref{fig:Temperature}(d), about 95\% of the time $\Delta T$ is 3 and 4, i.e., $\Delta T < 5^\circ$C, indicating that the Li-ion battery system is in good temperature uniformity.

The temperature and C-rate are two key factors that impact the aging rate of Li-ion batteries.
The aging speed can be described by aging rates, which are contained in the capacity fade curves.
Arrhenius's theory~\cite{J.Chem.Educ.61(1984)494K.J.Laidler, Z.Phys.Chem.4(1889)226S.Arrhenius, Z.Phys.Chem.4(1889)96S.Arrhenius} considering the effects of temperature and C-rate has been applied in the field of Li-ion batteries to determine activation barriers for aging~\cite{J.Electrochem.Soc.156(2009)A527Y.Zhang, J.PowerSources119(2003)874B.Y.Liaw} and other processes~\cite{J.PowerSources196(2011)5342J.P.Schmidt, Langmuir25(2009)12766Y.Yamada, J.PowerSources127(2004)72Z.Ogumi, J.Electrochem.Soc.151(2004)A1120T.Abe}.
A slope change in an Arrhenius plot indicates a mechanism change~\cite{J.PowerSources262(2014)129T.Waldmann}.
Arrhenius behavior has been observed in Li-ion batteries~\cite{J.PowerSources97-98(2001)13M.Broussely, J.PowerSources119(2003)874B.Y.Liaw, J.Electrochem.Soc.150(2003)A1385R.Bouchet}, showing a minimum of their aging rate at a certain temperature.
This minimum in the Arrhenius plot with a certain C-rate ($T = 5$--25$^\circ$C and $C_R < 0.2$ h$^{-1}$) points out the longest cycle life of Li-ion batteries~\cite{J.Electrochem.Soc.169(2022)030509M.Bozorgchenani, J.PowerSources262(2014)129T.Waldmann}.
For the discussed battery system, the aforementioned operation temperature and C-rate are close to the minimum in the corresponding Arrhenius plot, indicating a slower aging rate and longer life of the battery system.

\subsection{Daily Accumulated Capacity Distributions and Energy Usage of the Battery System}

Accumulated capacity (energy) is an important quantity for judging the health of batteries.
The accumulated capacity can be calculated from the source data by the trapezoidal rule, which was applied in Babylon before 50 BCE for integrating the velocity of Jupiter along the ecliptic~\cite{Science351(2016)482M.Ossendrijver}.
The trapezoidal rule works by approximating the region under the curve of the function $f(x)$ as a trapezoid and calculating its area.
In real cases, the composite trapezoidal rule with non-uniform spacing is used for accumulating data,
\begin{equation}
\int_{a}^{b} f(t)dt \approx \sum_{n=1}^{N}\frac{f(t_{n-1})+f(t_n)}{2}\Delta t_n,
\label{eq:CompositeTrapezoidalRule}
\end{equation}
where $t_n$ is a point in the partition of $[a, b]$ ($a = t_0 < t_1 < \cdots < t_{N-1} < t_N = b$), $\Delta t_n = t_n - t_{n-1}$ is the interval, and $n=0,1,2,...,N$ is an integer.
The accumulated capacity ($C_{acc}$) can be obtained by
\begin{equation}
C_{acc} = \int_{a}^{b} I(t)dt \approx \sum_{n=1}^{N}\frac{I(t_{n-1})+I(t_n)}{2}\Delta t_n,
\label{eq:AccumulatedCapacity}
\end{equation}
where $I(t_n)$ is the current value of the $t_n$ point from the source data.
The accumulated energy ($E_{acc}$) can be calculated by
\begin{equation}
\begin{split}
E_{acc} & = \int_{a}^{b} I(t)V(t)dt \\
        & \approx  \sum_{n=1}^{N}\frac{I(t_{n-1})V(t_{n-1})+I(t_n)V(t_n)}{2}\Delta t_n,
\end{split}
\label{eq:AccumulatedEnergy}
\end{equation}
where $V(t_n)$ is the total voltage value of the $t_n$ point from the source data.

\begin{figure}
\centering
\includegraphics[width=0.8\linewidth,keepaspectratio]{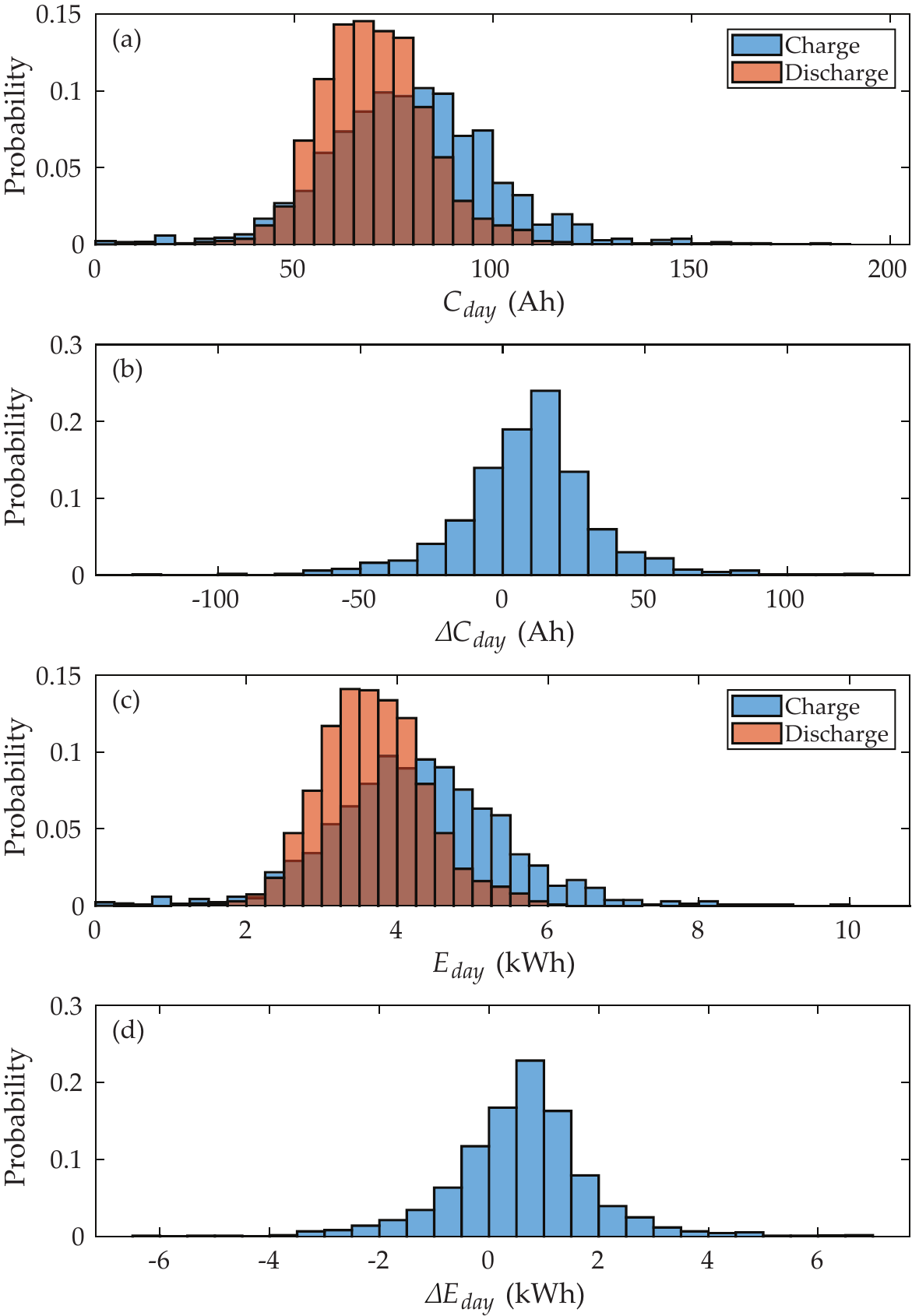}
\caption{
Daily accumulated capacity and energy distributions of the battery system.
(a) The distribution of accumulated charge and discharge capacities, where the accumulated charge and discharge capacities are marked by blue and red bars, respectively.
(b) The distribution of the daily accumulated capacity difference ($\Delta C_{day}$).
(c) The distribution of the daily accumulated charge and discharge energies.
(d) The distribution of the daily accumulated energy difference ($\Delta E_{day}$).
Time range of data: 2016-10-14$\sim$2020-07-20. (Total: 1376 days.)
}
\label{fig:Ah_kWh_histogram}
\end{figure}

Daily accumulated capacity and energy usages calculated by Eqs.~\ref{eq:AccumulatedCapacity} and~\ref{eq:AccumulatedEnergy} are shown in Fig.~\ref{fig:Ah_kWh_histogram}.
The major probability of the daily accumulated charge capacity ($C_{day}^{chg}$) distributes around 70--90 Ah, while the major probability of the daily accumulated discharge capacity ($C_{day}^{dis}$) is around 60--80 Ah (Fig.~\ref{fig:Ah_kWh_histogram}(a)).
The daily accumulated charge capacity is roughly higher than the daily accumulated discharge capacity, and such a feature is more obvious when we check the daily capacity loss, i.e., the difference between the daily accumulated charge and discharge capacities, $\Delta C_{day} = C_{day}^{chg} - C_{day}^{dis}$.
As shown in Fig.~\ref{fig:Ah_kWh_histogram}(b), $\Delta C_{day}$ exhibits a maximum probability of around 10--20 Ah.
The overall distribution of $\Delta C_{day}$ is located at the positive region, indicating that the daily accumulated input energy is higher than the daily accumulated output energy, i.e., an energy loss in the battery system.
The energy loss in the battery system can be explained by the equivalent circuit model (ECM), which is the most common battery model, consisting of resistance, capacitance, and inductance elements~\cite{J.PowerSources436(2019)226885J.Li, EnergyConvers.Manag.87(2014)367L.H.Saw, J.PowerSources174(2007)856M.Dubarry}.
The effective resistance causes joule heating during charging and discharging, resulting in energy loss in the battery system.

Daily accumulated energy as a counterpart to the daily accumulated capacity is depicted in Figs.~\ref{fig:Ah_kWh_histogram}(c) and (d).
Similar distributions are exhibited.
The major probability of the daily accumulated charge energy ($E_{day}^{chg}$) distributes around 3.75--4.75 kWh, while the major probability of the daily accumulated discharge energy ($E_{day}^{dis}$) is around 3.25--4.25 kWh (Fig.~\ref{fig:Ah_kWh_histogram}(a)).
The overall distribution of the difference between the daily accumulated charge and discharge energy ($\Delta E_{day} = E_{day}^{chg} - E_{day}^{dis}$) is located at the positive region with a value of about 0.5--1 kWh, indicating the feature of energy loss in the battery system.


The annual report about the accumulated capacity and energy usage of the battery system is listed in Table~\ref{tab:Ah_kWh_cycle}.
The annual accumulated charge capacity ($C_{year}^{chg}$) and annual accumulated discharge capacity ($C_{year}^{dis}$) exhibit the feature of $C_{year}^{chg} > C_{year}^{dis}$ under the scale in years.
The annual capacity loss rates ($CR_{year}^{loss}$) calculated by
\begin{equation}
CR_{year}^{loss} = 100\% \times \frac{(C_{year}^{chg} - C_{year}^{dis})}{C_{year}^{chg}}
\label{eq:AnnualCapacityLossRate}
\end{equation}
are of values from 8.2--13.9\%, and the capacity loss rate all time is about 11.2\%.
The cycle number per year can be calculated by discharge capacity per year over the nominal capacity of the battery system ($C_{year}^{dis} / C_{nom}$), where $C_{nom} = 250$ Ah is the nominal capacity of the discussed battery system.
To discuss energy usage explicitly, the annual accumulated charge energy ($E_{year}^{chg}$) and annual accumulated discharge energy ($E_{year}^{dis}$) are given.
The feature of $E_{year}^{chg} > E_{year}^{dis}$ under the scale in years remains.
The annual energy loss rates ($ER_{year}^{loss}$) calculated by
\begin{equation}
ER_{year}^{loss} = 100\% \times \frac{(E_{year}^{chg} - E_{year}^{dis})}{E_{year}^{chg}}
\label{eq:AnnualEnergyLossRate}
\end{equation}
are of values from 11.1--16.3\%, and the energy loss rate all time is about 13.8\%.




While the battery system is operating, it is impractical to obtain the full charge and discharge capacities.
Because we cannot arbitrarily shut down the system to do the CC-CV charge and CC discharge tests~\cite{Sci.Data8(2021)165H.C.Chung}.
The estimation of the battery system capacity fade becomes an open issue for the industry.
Experimental results demonstrated that at low C-rates ($C_R < 0.5$ h$^{-1}$), capacity fade is substantially affected by two parameters, i.e., cycle number and temperature, while the depth of discharge (DOD) has a negligible effect on capacity fade~\cite{J.PowerSources196(2011)3942J.Wang}.
For the discussed battery system, the average cycle number per year is about 103 and the total cycle number is 387 as listed in Table~\ref{tab:Ah_kWh_cycle}.
Compared to the data of LFP batteries (cycle life about 2000--4000 cycles with remaining capacity about 80\%)~\cite{BookChChung2021EngIntePotentialAppOutlooksLiIonBatteryIndustry, engrxiv2020OutlooksLiIonBatteriesH.C.Chung}, the battery system is far away from the end of life (EOL) with an estimated remaining capacity of more than 95\%.

\begin{table}
\centering
\caption{Annual capacity and energy usages of the battery system.
}
\begin{adjustwidth}{-\extralength}{0cm}
\begin{tabular}{c | c c c c | c c c}
\hline
Time interval & $C_{year}^{chg}$ (Ah) & $C_{year}^{dis}$ (Ah) & $CR_{year}^{loss}$ (\%) & Cycle number & $E_{year}^{chg}$ (kWh) & $E_{year}^{dis}$ (kWh) & $ER_{year}^{loss}$ (\%) \\
\hline
2016-10-14 $\sim$ 12-31 &   5458.6 &  4699.1 & 13.9 &  18.8 &  294.0 &  246.1 & 16.3 \\
                   2017 &  25434.2 & 23142.9 &  9.0 &  92.6 & 1371.8 & 1212.3 & 11.6 \\
                   2018 &  29690.8 & 26213.7 & 11.7 & 104.9 & 1606.0 & 1375.9 & 14.3 \\
                   2019 &  31587.8 & 27270.6 & 13.7 & 109.1 & 1706.3 & 1429.6 & 16.2 \\
2020-01-01 $\sim$ 07-20 &  16804.3 & 15422.9 &  8.2 &  61.7 &  908.0 &  807.1 & 11.1 \\
\hline
               All time & 108975.8 & 96749.2 & 11.2 & 387.0 & 5886.1 & 5071.0 & 13.8 \\
\hline
\end{tabular}
\end{adjustwidth}
\label{tab:Ah_kWh_cycle}
\end{table}

\subsection{Alpine Environmental Impacts on Battery Systems}

The impact of alpine environmental characteristics on battery systems can be evaluated from the perspective of two major physical quantities, i.e., temperature and pressure.
In terms of temperature, if the battery is placed outdoors, it may be too low, making it difficult to charge and discharge the battery.
Outdoors, there is also the problem of excessive temperature difference between day and night, which will cause battery life to be shortened.
If the battery is placed indoors, in a well-insulated environment, and combined with the heat generated by the operation of other equipment, the battery temperature may be moderate, and there will be no problem of excessive temperature difference between day and night. For example, the ESS in Paiyun Lodge is located indoors.
In terms of pressure, the air pressure in high mountains is low, which will cause expansion problems for soft-packed batteries.
Another point is the opening pressure of the battery safety relief valve. Since the valve opening pressure is designed for use on flat ground, moving to a mountain with lower outside air pressure, and the external pressure drops may cause the valve to open early.

\section{Electric Power Improvement Project for Paiyun Lodge}
\label{sec:ElectricPowerImprovementProject}

In this section, the status and problems of the old power system in Paiyun Lodge will be described first.
Then, some repair and optimization advice is listed.
The entire status of the electric power improvement engineering and other issues are stated at last.

\subsection{Status and Problems of the Old Power System in Paiyun Lodge}

\begin{figure}
\begin{adjustwidth}{-\extralength}{0cm}
\centering
\includegraphics[width=\linewidth,keepaspectratio]{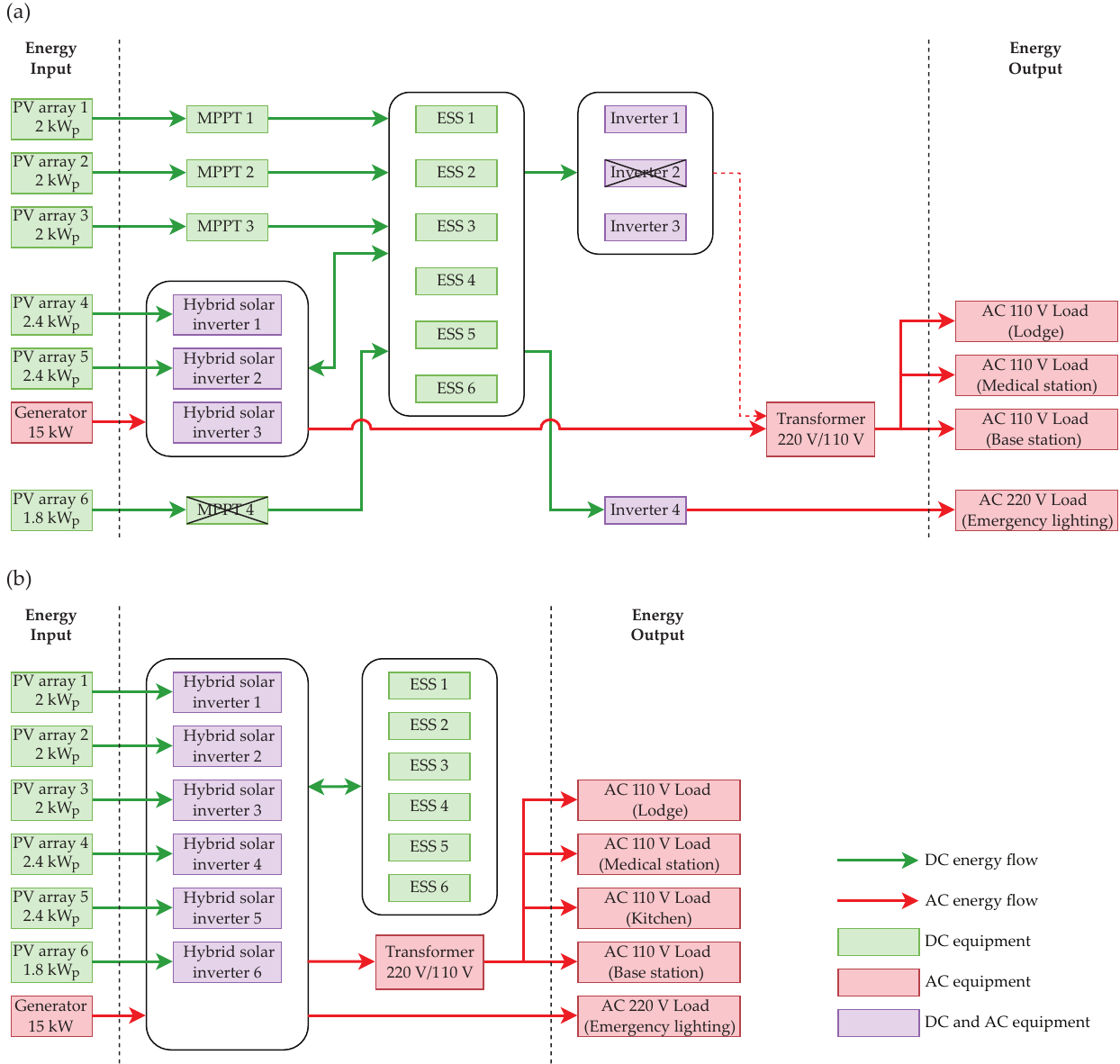}
\end{adjustwidth}
\caption{
Electricity architecture of the old and improved power systems in Paiyun Lodge.
(a) In the old power system, the DC power generated by the 1st, 2nd, 3rd, and 6th PV arrays is sent to the ESS by the 1st--4th MPPTs, and then converted to 220 V$_\mathrm{AC}$ power by the 1st--4th inverters. The transformer converts the AC power from 220 V$_\mathrm{AC}$ to 110 V$_\mathrm{AC}$ for general purposes (i.e., Paiyun Lodge, medical station, and base station). The 220 V$_\mathrm{AC}$ power for the emergency lighting is directly obtained from the 4th inverter. The rounded rectangles indicate that the equipment work in parallel. After 2016, the electricity for general purposes is provided by another power system, and the ESS is changed to Li-ion batteries from lead-acid batteries. The DC power generated by the 4th and 5th PV arrays is sent to the ESS by the 1st and 2nd hybrid solar inverters, and then converted to 220 V$_\mathrm{AC}$ power by the 1st--3rd hybrid solar inverters. The 15 kW generator serves as backup power for the system. The 2nd inverter and the 4th MPPT failed.
(b) In the improved power system, the DC power generated by the 6 PV arrays is sent to the ESS by the 6 hybrid solar inverters, and the converted 220 V$_\mathrm{AC}$ power is sent to the load for emergency lighting directly and sent to the load for general purposes by the transformer.
The electricity architecture and energy flow become simplified.
The green and red arrows indicate the DC and AC energy flow, respectively.
The green, red, and purple boxes indicate the DC, AC, and DC/AC equipment, respectively.
}
\label{fig:ElectricityArchitecture}
\end{figure}

The power architecture of the old system in Paiyun Lodge (containing energy creation, energy storage, and energy consumption) is shown in Fig.~\ref{fig:ElectricityArchitecture}(a).
The energy flows related to direct current (DC) and alternative current (AC) are colored in green and red, respectively.
To avoid confusion between DC and AC systems, the subscript of some units will add DC or AC to make the difference, e. g., V$_\mathrm{DC}$ indicates DC voltage.
In energy creation, the 1st, 2nd, 3rd, and 6th PV arrays with a total solar capacity of 7.8 kW$_\mathrm{p}$ were installed before 2016. After passing through the maximum power point trackers (MPPTs), the power directly enters the Li-ion battery ESS. The 4th and 5th PV arrays with a total solar capacity of 4.8 kW$_\mathrm{p}$ were installed after 2016, and their power is passed to the ESS by hybrid solar inverters. A diesel generator is used as a backup and emergency power source for the hybrid solar inverters.
In energy storage, it is provided by a 48 V$_\mathrm{DC}$ Li-ion battery ESS (composed of 6 battery cabinets with a total capacity of 72 kWh).
In energy consumption, the 220 V$_\mathrm{AC}$ power is mainly provided by the hybrid solar inverters, which is converted into 110 V$_\mathrm{AC}$ power by the step-down device, and then supplies power to the loads of the Paiyun Lodge, the medical station, and the base station.

\begin{figure}
\centering
\includegraphics[width=0.6\linewidth,keepaspectratio]{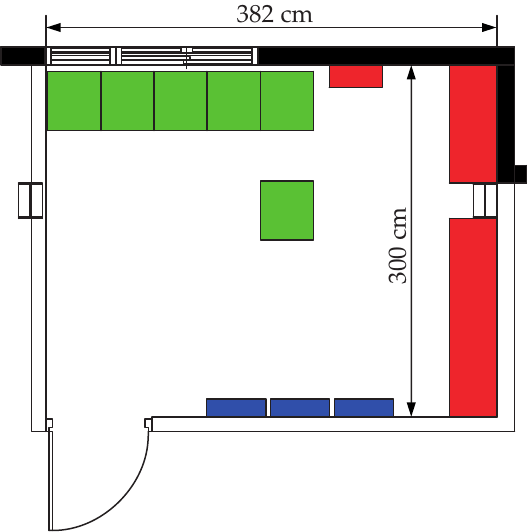}
\caption{
Map of the machine room (unit: cm).
The 48V$_\mathrm{DC}$ Li-ion battery ESS composed of 6 Li-ion battery cabinets with a total energy of 72 kWh is indicated by green boxes.
Three hybrid solar inverters on the wall with a maximum output power of 15 kW are indicated by blue boxes.
Seven distribution boxes on the wall connecting energy input and output systems are indicated by red boxes.
}
\label{fig:SiteMap}
\end{figure}

On the roof of the lodge are the 1st--6th PV arrays with solar capacity of 2 kW$_\mathrm{p}$, 2 kW$_\mathrm{p}$, 2 kW$_\mathrm{p}$, 2.4 kW$_\mathrm{p}$, 2.4 kW$_\mathrm{p}$, and 1.8 kW$_\mathrm{p}$, respectively.
The major equipment and distribution boxes locate in the machine room as shown in Fig.~\ref{fig:SiteMap}.
The ESS composed of 6 Li-ion battery cabinets is indicated by boxes in green colors.
Each Li-ion battery cabinet is made of a battery pack with a nominal voltage of 48 V$_\mathrm{DC}$ and a BMS.
The 48 V$_\mathrm{DC}$ battery pack is composed of LFP batteries with a 16S5P configuration, in other words, there are 5 battery cells connected in parallel as a group and 16 groups of battery cells connected in series as a whole battery pack of 80 battery cells.
Each battery cells possess a capacity of 50 Ah, and the capacity of each cabinet is about 12 kWh.
The entire capacity of the ESS is about 72 kWh.
Three hybrid solar inverters are on the wall with a maximum output power of 15 kW (blue boxes in Fig.~\ref{fig:SiteMap}).
Seven distribution boxes are on the wall, connecting the PV arrays, hybrid solar inverters, old MPPTs, old inverters, generator, transformer and loads (red boxes in Fig.~\ref{fig:SiteMap}).

Several problems of the old power system are listed below.
\begin{enumerate}
\item \textbf{Long-time operation causes system aging.}
The off-grid PV ESS of Paiyun Lodge was installed in Oct. 2016, and it has entered its seventh year of operation so far. In the past seven years, the system has been working 24 hours a day without interruption, and the aging of equipment is inevitable. In particular, the aging of the semiconductor components in the inverter is more obvious, which has gradually caused the system to stop without warning and from time to time. It must be restarted before the system can be used again. This is a warning sign that the hybrid solar inverters need to be replaced to avoid the system going further into a more serious condition. In addition, once a fault occurs, the system will be shut down for a long time. Whether it is an MPPT or a hybrid solar inverter, it has been more than 6 years. Among them, the MPPT was installed 17 years ago. These devices are already discontinued products. If they fail to repair, it is a difficult and time-consuming work. It might takes several months to half a year to get the devices from the manufacturer, which will make the system shut down for a long time.
\item \textbf{Imbalance of battery voltages in the energy storage cabinet and some batteries are faulty.}
After seven years of operation, the Li-ion batteries have been in a state of imbalanced battery voltages. In the past few years, it has been found that the situation has become more and more serious. In addition, some batteries have been in serious bad condition and are judged to be faulty, which has seriously affected the overall operation of the system.
\item \textbf{Power paths to the ESS are too complex.}
The power path complexity of the ESS is lifted by two kinds of power paths.
The ``charge and discharge through different ports'' is used for the power input of 1st--4th MPPTs and the power output of 1st--3rd inverters.
The ``charge and discharge through the same port'' is used for the 1st--3rd hybrid solar inverters.
When installing another power system in 2016, the power paths of the two systems are combined in the Li-ion battery ESS to enlarge the solar power capacity.
This situation increases the power path complexity of the ESS, with a consequent increase in the probability of system failure.
\item \textbf{Solar idle for older systems.}
The power of the 6th PV array cannot be used due to the failure of the 4th MPPT (as shown in Fig.~\ref{fig:ElectricityArchitecture}(a)), and the solar energy of 1.8 kW$_\mathrm{p}$ is wasted.
\end{enumerate}

\subsection{Repair and Optimization Advice to the Government}

A proposal about the repair and optimization advice is made in 2021.
Some points are listed below.
\begin{enumerate}
\item \textbf{Replace hybrid solar inverters and integrate PV power paths.} The power of all PV arrays directly passes through the hybrid solar inverters, and then goes to the Li-ion battery ESS. By replacing the hybrid solar inverter with a new version, the maintenance problem of the system can be solved and the system can be prevented from being shut down for a long time. In addition, the power of the 6th PV array could be used again by the new hybrid solar inverter, the power can be reconnected into the lodge for use, increasing the energy supply of the lodge.
\item \textbf{Reduce the power path complexity of the ESS.} By replacing the hybrid solar inverters, two kinds of power paths will become one (i.e., ``charge and discharge in the same port''), reducing the complexity of the power path and failure possibility.
\item \textbf{Replace new batteries and install active voltage balancer.} Replacing the faulty Li-ion batteries to solve the influence of the bad Li-ion battery ESS. On the other hand, active balancers are installed for the ESS to actively balance the voltage of the unbalanced batteries, prolong the service life of the batteries, and improve the system stability.
\item \textbf{Construct cloud EMS.} Monitor the energy input, output and battery conditions of the entire off-grid PV ESS. The relevant data is stored in the cloud database, so that remote engineers can know the real-time status of the system, and pre-process before the system has obvious problems to avoid serious incidents, Maintaining system health and improving system usage. At the same time, the EMS can also provide statistical data of energy use, annual total power generation, annual total power consumption, etc., for the competent unit to evaluate the energy usage status.
\end{enumerate}

\subsection{Engineering Status of the Power Improvement Project}

\begin{figure}
\centering
\includegraphics[width=0.8\linewidth,keepaspectratio]{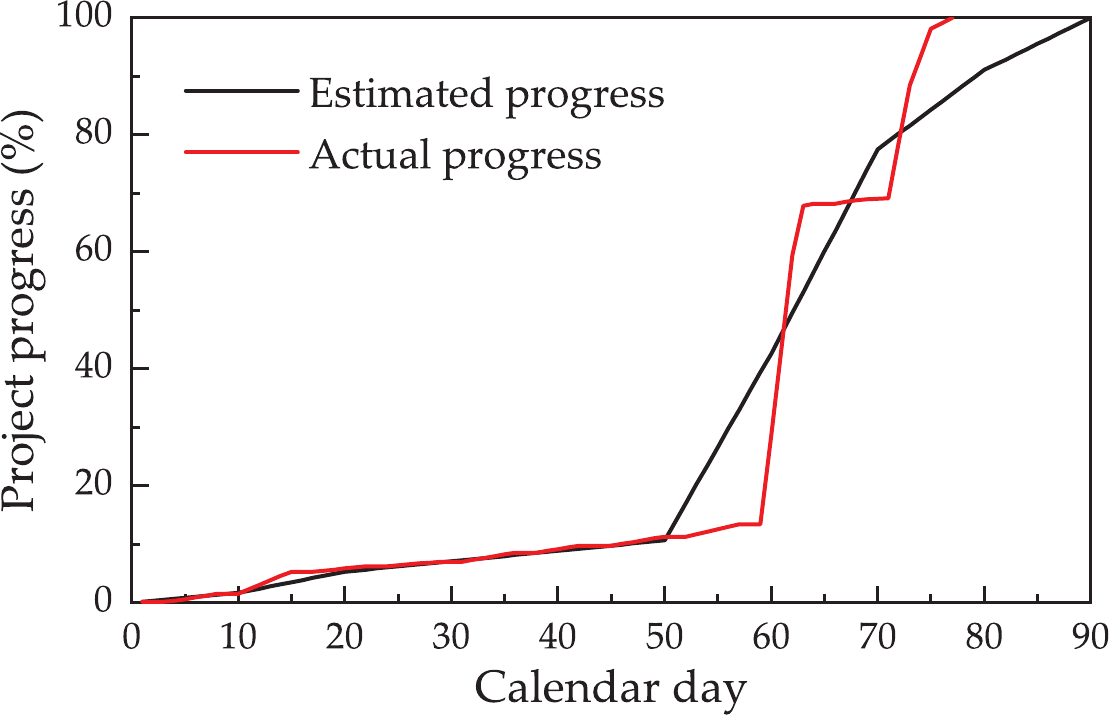}
\caption{
S-curve of the project.
The estimated progress and actual progress are indicated by the black and red curves, respectively.
The project was completed 13 calendar days ahead of schedule.
}
\label{fig:EngineeringScurve}
\end{figure}

The engineering project started on Jul. 29th, 2022, and the scheduled completion date is Oct. 26th, 2022.
The approved construction period is 90 calendar days.
The project was completed on the 77th calendar day of the cumulative construction period (13 calendar days in advance) as shown in Fig.~\ref{fig:EngineeringScurve}.
The project status is described below.

\begin{figure}
\centering
\includegraphics[width=0.8\linewidth,keepaspectratio]{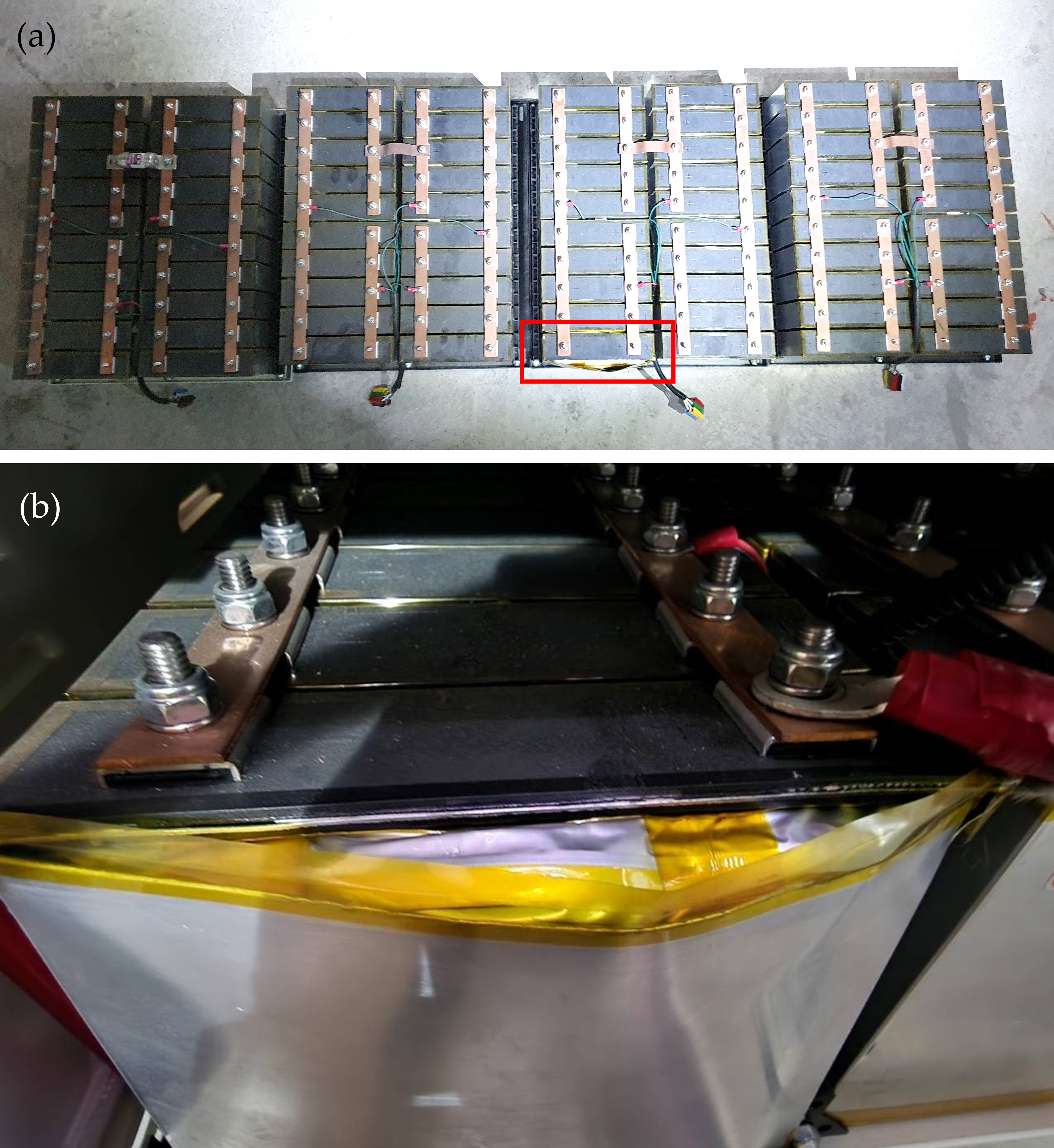}
\caption{
(a) Remove all battery modules from the battery cabinet, and check the appearance and voltages. Taking this energy storage cabinet as an example, there are 16 strings of battery cells, each string is connected in parallel with 5 battery cells, for a total of 80 battery cells.
Swollen Battery Cells are marked by the red rectangle.
(b) Swollen Battery Cells should be replaced.
}
\label{fig:ReplacingSwollenBatteryCells}
\end{figure}

\textbf{Reform and optimization of existing Li-ion battery cabinets.}
Replacing of old and faulty batteries and installation of active balancers are included.
In the replacement of old and faulty battery cells, if the appearance is obviously swollen or there is leakage or voltage in an abnormal range, the battery should be replaced.
Each 48 V$_\mathrm{DC}$ energy storage cabinet is made by a 16S5P configuration.
All battery modules were removed from the battery cabinet for checking as shown in Fig.~\ref{fig:ReplacingSwollenBatteryCells}.
The 4th energy storage cabinet has 3 groups of battery cells that have swelled significantly and abnormal voltages of 1.000 V$_\mathrm{DC}$ and 0.967 V$_\mathrm{DC}$.
(For an LFP battery cell, the normal operation voltage is about 2.5--3.65 V~\cite{Sci.Data8(2021)165H.C.Chung, BookChChung2021UL1974}.)
Each group of battery cells needs to be replaced with 5 battery cells, for a total of 15 cells.
The 5th energy storage cabinet has 2 groups of battery cells that are obviously swollen, and each group of battery cells needs to be replaced with 5 battery cells, for a total of 10 cells.
All replaced a total of 25 batteries.
The LFP battery cell with a capacity of 50 Ah is used in the replacement, which is made by Chang Hong Energy Technology Co., Ltd., Taiwan.

Installation of active balancers to solve the problem of large voltage differences among battery groups.
The active balancer with a maximum balancing current of 10 A is used.
Compared with a 250 Ah battery, the maximum charge and discharge current of 10 A corresponds to $C_R = 0.04$ h$^{-1}$, which means that the balance current should be used to balance the voltage, without causing a large voltage drop or rise.
During installation, measuring the balance current between two adjacent battery groups is one of the important checkpoints.
A non-zero balance current indicates that the active balancer works well.
Since the balance current is small, it requires 1 to 2 days to observe the condition after installing the active balancers.
The average voltage differences ($\Delta V_{avg}$) during charging for all energy storage cabinets are dropped below 60 mV (as shown in Table~\ref{tab:AverageDeltaV}).
$\Delta V_{avg}$ shrinkage is an effective way to enlarge the available capacity of the old Li-ion battery ESS.
Furthermore, voltage balancers can effectively avoid overcharging and overdischarging the battery, effectively prolonging the battery life~\cite{PowerElectronics14(2016)24Y.K.Chen}.
As one energy cabinet is under reform and optimization, the rest 5 cabinets still operate and provide the lodge with uninterrupted electricity.

\begin{table}
\centering
\caption{Average voltage differences $\Delta V_{avg}$ during charging before and after installation active balancers.
With active balancers, $\Delta V_{avg}$s are dropped below 60 mV.
}
\begin{tabular}{c c c}
\hline
Cabinet No. & $\Delta V_{avg}$ (mV) & $\Delta V_{avg}$ (mV) \\
 & without balancers & with balancers \\
\hline
1 & 163 & 38 \\
2 & 124 & 57 \\
3 & 153 & 20 \\
4 & 2552$^*$ & 37 \\
5 & 152 & 21 \\
6 & 34 & 26 \\
\hline
\end{tabular}
\begin{flushleft}
$^*$ The value of $\Delta V_{avg}$ is very large due to the low voltages of the faulty battery cells.
\end{flushleft}
\label{tab:AverageDeltaV}
\end{table}

\textbf{Solar inverter system reformation and optimization.}
Four MPPTs, four inverters, and three hybrid solar inverters of the old power system are dismantled (Fig.~\ref{fig:ElectricityArchitecture}(a)).
In the improved system, six new-version hybrid solar inverters are installed as shown in Fig.~\ref{fig:ElectricityArchitecture}(b).
The electricity of the six PV arrays enters the six hybrid solar inverters, respectively.
All inverters will operate in parallel mode and output one-phase-two-wire (1P2W) 220 V$_\mathrm{AC}$ power at 60 Hz.
The 220 V$_\mathrm{AC}$ power is sent to the load for emergency lighting directly and sent to the load for general purposes through the 220 V$_\mathrm{AC}$/110 V$_\mathrm{AC}$ transformer.
Because the maintenance and protection on high mountains is not easy, it will be handled by over-installation.
Six hybrid solar inverters can disperse the load of the lodge.
As one inverter fails, the other five can still operate, and the load of the survival inverters will increase by about 20\% each.
If this happens on the old system, the increasing load is about 50\%.
Hence, the over-installation can effectively improve the working stability of the system.
In addition, it's time-consuming work to install 6 hybrid solar inverters at once.
There is a skill for reducing outage time.
In the early stage, We can install two hybrid solar inverters to provide temporary electricity.
After the rest four hybrid solar inverters are installed on the wall, all inverters can be back to line at once.

\textbf{Reorganization of distribution boxes and power line adjustment.}
The power lines of old and abandoned systems have been removed for safety.
Distribution boxes as power connections and breakpoints have been reorganized and the number of distribution boxes reduces from 9 to 7, indicating that the power lines become simpler.
Four distribution boxes connect hybrid solar inverters directly, i.e., one controlling the input power from PV arrays, one controlling the input/output power from the ESS, one controlling the 220 V$_\mathrm{AC}$ input power (power from the generator), and one controlling the 220 V$_\mathrm{AC}$ output power.
Such an organization can physically separate hybrid solar inverters from PV arrays, the ESS, the generator, and the AC power output.
If one or several inverters failed, the failed part can be physically separated for fixing and the rest part of the system can still operate and provide electricity.
There are 3 distribution boxes for power output.
One is for the transformer converting power from 220 V$_\mathrm{AC}$ to 110 V$_\mathrm{AC}$.
One controls the 110 V$_\mathrm{AC}$ power to general-purpose usages, such as the lodge, medical station, kitchen, and base station.
One controls the 220 V$_\mathrm{AC}$ power to emergency lighting.
On the other hand, after reorganizing the power lines of the PV arrays, the 6th PV array is back to the line, where a 16.7\% solar capacity is recovered.
In addition, it is safe to adjust the power lines of PV arrays at night without affecting the power operation of the lodge.



\begin{figure}
\begin{adjustwidth}{-\extralength}{0cm}
\centering
\includegraphics[width=\linewidth,keepaspectratio]{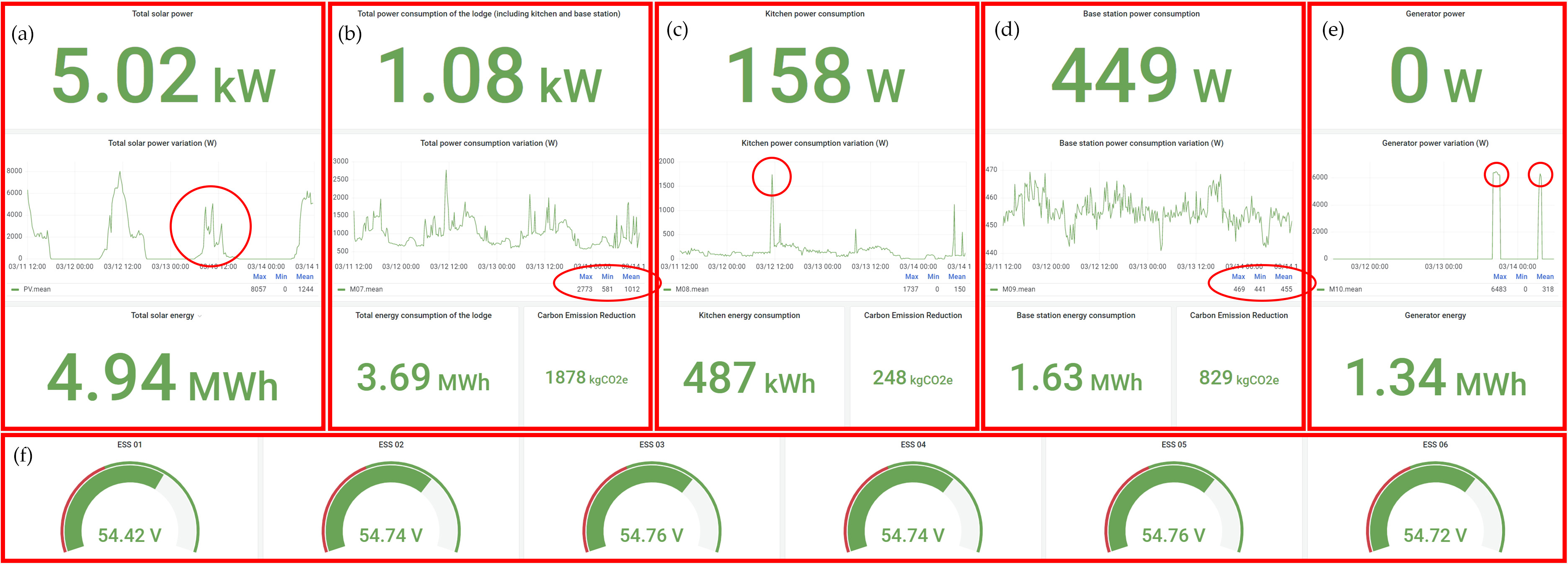}
\end{adjustwidth}
\caption{
User monitor screen of the cloud EMS.
The screen can be divided into five zones (red rectangles) for realizing the energy information.
(a) Energy generation information of all PV arrays, including total solar power, total solar power variation, and total solar energy.
(b)--(d) Energy consumption information of the (b) Paiyun Lodge, (c) kitchen, and (d) base station, including total power consumption of the lodge, total power consumption variation (with maximum, minimum, and mean values), total energy consumption, and carbon emission reduction.
(e) Energy generation information of the diesel generator.
(f) Total voltages of 6 energy storage cabinets.
}
\label{fig:CloudEMS}
\end{figure}

\textbf{Installation of cloud EMS.}
The self-developed cloud EMS~\cite{BookChChung2021EngIntePotentialAppOutlooksLiIonBatteryIndustry, engrxiv2020OutlooksLiIonBatteriesH.C.Chung} is installed to remotely monitor the off-grid PV ESS.
When there is any situation in the system or the aging of the battery pack begins to worsen, it is convenient for the crew to repair and maintain the system immediately, maintain the health of the system, and improve the utilization rate of the system.
The energy input (from 6 PV arrays and the diesel generator) and energy output (to the Paiyun Lodge, kitchen, and base station) are monitored (as shown in Fig.~\ref{fig:CloudEMS}).
The total voltages of all energy cabinets are also listed for reference.
The cloud server is set up to store energy information in the database for more than 10 years.
Its security stability should have the ability to block some basic hacker attacks, e.g., distributed denial of service (DDoS) attacks~\cite{Computer50(2017)80C.Kolias, J.Netw.Comput.Appl.67(2016)147O.Osanaiye, ACMComput.Surv.39(2007)3T.Peng}.
A backup database server at the local site is also set for data redundancy.
Moreover, the EMS can also provide statistical energy data (such as annual total energy generation, annual total energy consumption, etc.) for the competent unit to evaluate the energy usage status.



Many conditions of energy usage can be read from the monitor screen (Fig.~\ref{fig:CloudEMS}).
The current energy input and output are listed on the top row.
For example, the current PV power, power consumption of the lodge, kitchen, and base station, as well as diesel generator power are 5.02 kW, 1.08 kW, 158 W, 449 W, and 0 W, respectively.
The historical energy input and output data stored in the database can provide detailed energy analysis.
For example, within three days, the power variation curve of all PV arrays exhibited a non-smooth curve with many sharp power drops (circle in Fig.~\ref{fig:CloudEMS}(a), reflecting the feature of fluctuations in renewable energies~\cite{IEEETrans.SmartGrid6(2015)124K.Rahbar, IEEETrans.Sustain.Energy1(2010)117S.Teleke}.
The power consumption of the lodge demonstrated a maximum value of 2773 W, a minimum value of 581 W, and a mean value of 1012 W (circle in Fig.~\ref{fig:CloudEMS}(b)).
The major power peaks of the lodge were contributed by the kitchen (circle in Fig.~\ref{fig:CloudEMS}(c)).
At the same time, it was verified that high-power appliances (about 1500 W) are used in the kitchen.
The power consumption of the base station exhibited a steady value of 455$\pm$14 W with an energy fluctuation of 3.1\% (circle in Fig.~\ref{fig:CloudEMS}(d)).
Such fluctuation indicated that no explicit dependence was found.
The diesel generator had been operated for a while with a peak power of about 6400 W to the ESS (circles in Fig.~\ref{fig:CloudEMS}(e)).

\textbf{Alpine worker carrying.}
Since 2013, alpine worker carrying becomes the only way to take cargo up the mountain because of the tragedy of the helicopter crash at Mt. Jade North Peak~\cite{LibertyTimes(2013)ThreeKilledYushanHelicopterCrash, LibertyTimes(2014)ReasonsYushanHelicopterCrash}.
From Mt. Jade Trailhead at Tataka Saddle (at an altitude of 2610 m (8563 ft)) to Paiyun Lodge (at an altitude of 3402 m (11161 ft)), the distance is about 8.5 km (5.3 mi) with an altitude rise of 800 m (2625 ft).
The handling weight of the Alpine worker is generally 30--60 kg.
Tools and equipment must be disassembled and packed into packages of less than 30 kg for easy handling.
This project transports about 300 kg of tools and equipment up the mountain by alpine workers.




\textbf{Demolition and transportation of old and abandoned facilities.}
For maintaining the natural environment on the high mountain, the old and abandoned facilities should be demolished and transported down the mountain in line with the principle of Leave No Trace (LNT)~\cite{The7PrinciplesLeaveNoTrace(LNT)}.
This project transports about 300 kg of old and abandoned facilities down the mountain by alpine workers.




\textbf{Material preparation, occupational safety, quality control, safety and health facilities, etc.}
(1) Materials for the project are sometimes not readily available. It must be prepared in advance to avoid unforeseen and inability to carry out construction within the project period.
(2) Occupational safety is very important during engineering.
Engineers must not only have knowledge of safe engineering, but also have relevant protective equipment to protect themselves, such as safety helmets, safety glasses, masks, and electrical insulation equipment.
(3) Quality control is not only in material acquisition, but also in construction quality. Appropriate checkpoints can improve construction quality and construction efficiency.
One of the ways to improve the quality of construction is to cooperate with the opinions of the supervision unit to carry out construction.
(4) Some local engineering code requirements must also be met.
For example, ``construction notification board'' and ``occupational safety and health board'' should be set up on site.
Engineers require professional licenses to meet local regulations, such as ``Safety and health education and training for manager of occupational safety and health affairs.''

\textbf{Other engineering issues.}
\textbf{(1) Establish the kitchen power line and monitor the energy usage.}
During this construction, the staff responded that the kitchen may use high-power electrical appliances such as electric cookers, which made the lodge have doubts about the safety of electricity.
After all, when the off-grid PV ESS was installed in the past, its energy capacity was only considered for the supply of lighting and medical equipment in the lodge, and it was not planned for the use of high-power electrical appliances in the kitchen. Although the total power of the hybrid solar inverters can supply the high-power load, the power lines in the kitchen are pulled from elsewhere using extension cords, which were designed for low-power light-emitting diode (LED) lighting equipment. Under high power conditions, there may be a public security problem that the wire is overloaded and melted.
In order to solve this problem, after discussing with the chief, an individual power line was pulled from the 110 V$_\mathrm{AC}$ distribution box for the exclusive use of the kitchen. At the same time, the power monitoring of the individual power line was also carried out. It can be seen from the cloud EMS that high-power appliances are indeed used in the kitchen before the meal period (as shown in Fig.~\ref{fig:CloudEMS}(c)).
\textbf{(2) Insufficient energy capacity of the ESS.}
Since the energy consumption of the base station and kitchen is not considered during the design stage of the PV ESS, the energy capacity of the ESS is not sufficient, especially for the aging system.
Although, in this project, some new techniques have been applied to extend the life cycle of the system, the lack of energy capacity remains a problem.
It is recommended to expand the energy capacity in the future.



\begin{table}
\caption{Differences between before and after the improvement of the power system.}
\begin{adjustwidth}{-\extralength}{0cm}
\centering
\begin{tabular}{c p{228pt} p{228pt}}
\hline
No. & Before improvement & After improvement \\
\hline
\rowcolor{LightCyan}
1 & The power system is old and has irregular failures. Once a serious failure, the repair time will be long, and the system will be down for a long time. Whether it is an MPPT or a hybrid solar inverter, it has been more than 6 years. Among them, the MPPTs were installed almost two decades ago. These devices are already discontinued products. If they fail to repair, it is not an easy task. Even if the manufacturer provides the products, it might still take several months to half a year to obtain the products, which will make the system shut down for a long time. & Replacing the new version of the hybrid solar inverters can solve the maintenance problem of the system and prevent the system from shutting down for a long time. Over-installation of the hybrid solar inverters is a good way for system maintenance and protection on high mountains.\\
2 & Power paths to the ESS are too complex. Two kinds of power paths will increase the probability of failure. & The Power path to the ESS becomes simple. One kind of power path is used, reducing the failure probability.\\
\rowcolor{LightCyan}
3 & The power of the 6th PV array cannot be used due to the failure of the 4th MPPT. & Through the new version of the hybrid solar inverter, the power of the 6th PV array can be reconnected to the lodge for use, increasing the energy supply of the Paiyun Lodge. About 16.7\% solar capacity is recovered.\\
4 & The load is supplied by 3 hybrid solar inverters. The load of each inverter is relatively large, and the service life of the machine will be shorter. & The load is supplied by 6 hybrid solar inverters. The load of each inverter is relatively small, and the service life of the machine will be longer.\\
\rowcolor{LightCyan}
5 & Since there are only 3 hybrid solar inverters, when one fails, the load of the other two will increase by 50\%, which is relatively dangerous and may not continue to supply power to the lodge. & Since there are 6 hybrid solar inverters, when one of the inverters fails, the load of the other five will increase by 20\%, which is relatively safe. The system can continue to operate the lodge when one of the inverters fails.\\
6 & The generator transmits power to 3 hybrid solar inverters, and the maximum charging efficiency for the ESS is poor. & The generator transmits power to 6 hybrid solar inverters, and the maximum charging efficiency for the ESS is better.\\
\rowcolor{LightCyan}
7 & 9 distribution boxes, full of lines of different systems, complicated lines,  difficult maintenance. & 7 distribution boxes, simple wiring, easy maintenance.\\
8 & Some batteries are faulty, causing operation problems in the system. & Replacing new batteries and bringing the system back to work normally.\\
\rowcolor{LightCyan}
9 & Imbalance of battery voltages in the energy storage cabinet. & The active balancers are installed to balance the battery voltages.\\
10 & No EMS. & The cloud EMS is installed and can monitor the system remotely. When there is any situation in the system or the aging of the battery pack begins to worsen, it is convenient for the crew to repair and maintain the system immediately, maintain the health of the system, and improve the utilization rate of the system. At the same time, the EMS can also provide statistical data on energy use, annual total power generation, annual total power consumption, etc., for the competent unit to evaluate the energy use status.\\
\hline
\end{tabular}
\end{adjustwidth}
\label{tab:DifferenceBeforeAfterImprovement}
\end{table}

\textbf{Risks and obstacles.}
Some obstacles to this alpine engineering project are listed.
\textbf{(1) Acute mountain sickness (AMS).}
AMS, caused by rapid exposure to low amounts of oxygen at a place of high altitude, is the most common form of acute altitude illness and typically happens in unacclimatized persons ascending to altitudes $>$ 2500 m (or $>$ 8000 ft)~\cite{HighAlt.Med.Biol.19(2018)4R.C.Roach, HighAlt.Med.Biol.20(2019)192M.Burtscher, Ann.Intern.Med.118(1993)587B.Honigman, Lancet308(1976)1149P.H.Hackett, N.Engl.J.Med.280(1969)175I.Singh}.
People can respond to high altitude with various symptoms, such as headache, nausea, vomiting, fatigue, tiredness, confusion, trouble sleeping, and dizziness~\cite{HighAlt.Med.Biol.19(2018)193M.Horiuchi, PlosOne12(2017)e0183207F.Y.Cheng, WildernessEnviron.Med.27(2016)371A.N.Issa, HighAlt.Med.Biol.15(2014)446M.Burtscher, Mil.Med.132(1967)585J.E.Hansen}.
The incidence of AMS may be accompanied by severe or even fatal high-altitude pulmonary edema (HAPE)~\cite{ExpertOpin.Pharmacother.19(2018)1891K.E.Joyce, Compr.Physiol.2(2012)2753E.R.Swenson, J.Appl.Physiol.98(2005)1101P.Bartsch, Circulation103(2001)2078M.Maggiorini, Lancet1(1965)229I.Singh, Medicine40(1961)289H.N.Hultgren, N.Engl.J.Med.263(1960)478C.S.Houston} or high-altitude cerebral edema (HACE)~\cite{J.Appl.Physiol.131(2021)313R.E.F.Turner, AJNRAm.J.Neuroradiol.40(2019)464P.H.Hackett, Cell.Mol.LifeSci.66(2009)3583D.M.Bailey, HighAlt.Med.Biol.5(2004)136P.H.Hackett, WildernessEnviron.Med.11(2000)89B.Basnyat, J.Am.Med.Assoc.280(1998)1920P.H.Hackett}.
The Lake Louise scoring system (LLSS) was designed to evaluate adults for symptoms of AMS~\cite{HighAlt.Med.Biol.19(2018)4R.C.Roach, HighAlt.Med.Biol.8(2007)124A.Southard,  Aviat.SpaceEnviron.Med.66(1995)963G.Savourey}.
Up to now, it is the most commonly used scoring system used to assess AMS, which was introduced in 1991 and was last modified in 2018~\cite{HighAlt.Med.Biol.19(2018)4R.C.Roach}.
LLSS uses an assessment questionnaire to determine whether an individual has no AMS, mild AMS, or severe AMS.
During climbing or at high altitudes, LLSS is an appropriate tool to evaluate our health.
\textbf{(2) Risks of climbing high mountains.}
High-altitude climbing has many risks, engineers (who are also climbers) should be prepared for better injury prevention and reducing search-and-rescue events~\cite{Clin.Transplant.29(2015)1013K.S.Suh, WildernessEnviron.Med.24(2013)221R.C.Mason}, e.g., maintaining adequate physical strength and adequate equipment.
\textbf{(3) Weather, accidents and natural hazards.}
Low temperatures, strong wind, and rain will make difficult climbing and affect engineering.
Typhoons~\cite{Nat.Geosci.9(2016)753W.Mei, Mon.WeatherRev.131(2003)1323C.S.Chen, Terr.AtmosphericOcean.Sci.33(2000)26G.J.Jian, Bull.Am.Meteorol.Soc.80(1999)67C.C.Wu} and earthquakes~\cite{Terr.Atmospheric.Ocean.Sci.33(2022)24K.C.Chen, Eng.Geol.71(2004)49C.W.Lin, Geology32(2004)733S.J.Dadson, Tunn.Undergr.SpaceTechnol.16(2001)133W.L.Wang, Science288(2000)2346H.Kao} are common natural disasters in Taiwan. In severe situations, work stoppages or evacuation may be required.
During the project period, a severe 6.9-magnitude earthquake occurred on Sep. 18th~\cite{Guardian918Earthquake, InitiumMedia(2022)1Dead146Injured918Earthquake, PTS(2022)QuakeDamagesAbout370HualienHomes}, affecting the progress of the project.
Including foreshocks, main shocks, and aftershocks, there are about 200 earthquakes in this earthquake sequence~\cite{AppleNews(2022)918AftershocksIncreased, PTS(2022)EarthquakeHitTaitung, LTN(2022)AftershocksContinue, PTS(2022)SleepingOutsideAfterEarthquake}.
When in Paiyun Lodge, we also encountered several felt earthquakes.
\textbf{(4) Lack of equipment supply.}
There are no convenience stores or hardware stores in the high mountains. If we forget to bring tools and equipment, the relevant section can only be abandoned, and the construction can only be carried out after the next time we go up the mountain.
In particular, for projects going to Mt. Jade for several days, we must first apply to the Yushan National Park Headquarters. Administrative procedures can also increase time consumption.

\begin{figure}
\centering
\includegraphics[width=\linewidth,keepaspectratio]{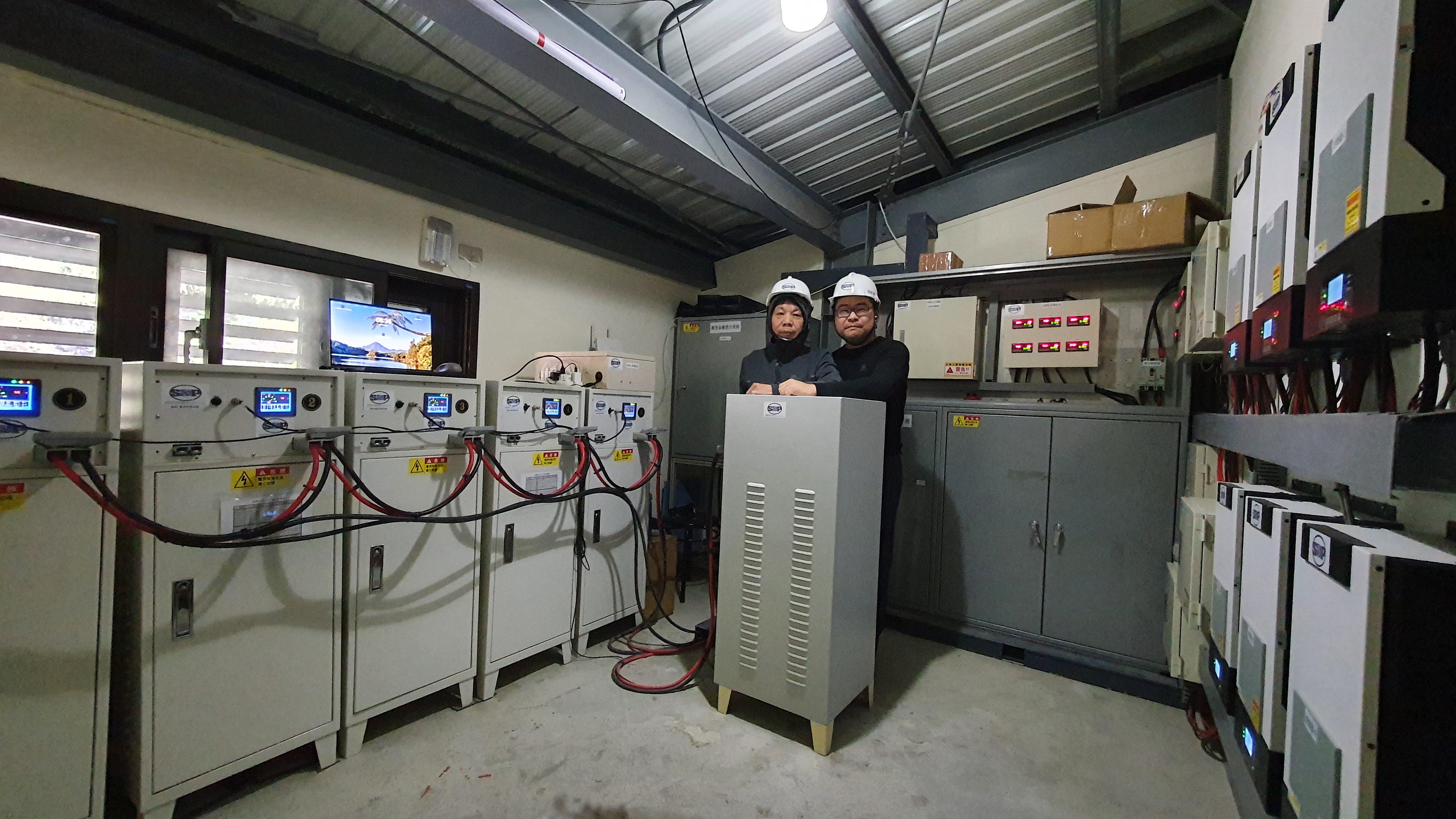}
\caption{
Site photo after the completion of the off-grid PV ESS improvement (The photo was taken on Oct. 10th, 2022).
The ESS consisting of 6 energy storage cabinets is on the left side, and the hybrid solar inverters with major distribution boxes are on the right side.
The people standing in the middle are Jung-Feng Jack Lin (left) and Hsien-Ching Chung (right).
}
\label{fig:FinishPhoto}
\end{figure}

\textbf{The improved power system started operation.}
The project was completed on Oct. 13th, 2022, and the off-grid PV ESS started trial operation (as shown in Fig.~\ref{fig:FinishPhoto}).
The completion confirmation meeting was held on Oct. 21st, 2022.
On Nov. 10th, 2022, the project acceptance procedure was carried out, and the system was officially operating.

\section{Simple Cost Analysis Between Medium-Scale Lead-Acid and Li-Ion Battery Systems}
\label{sec:BriefCostAnalysis}

Based on the experience of this power improvement project, a brief cost model can be established for analysis of the usage cost between medium-scale lead-acid and Li-ion battery systems.
The total cost ($Cost$) consists of two parts
\begin{equation}
Cost = Cost_B + Cost_M,
\label{eq:CostAnalysis}
\end{equation}
where $Cost_B$ is the battery cost and $Cost_M$ is the extra moving cost due to high mountains.
\begin{equation}
Cost_B = Price_B \times Energy_B,
\label{eq:CostBattery}
\end{equation}
\begin{equation}
Cost_M = Price_M \times ( 2 \times Energy_B / D_E ),
\label{eq:CostMoving}
\end{equation}
where $Price_B$ is the battery price per kilogram, $Energy_B$ is the total energy of the battery system in unit of kWh, $Price_M$ is the moving price per kilogram, $D_E$ is the energy density of the battery system in unit of kWh/kg, as well as ``2'' indicates uphill and downhill.
Before calculation, some assumptions should be noted.
Assumption 1: The hybrid inverter can normally work with lead-acid and Li-ion batteries, avoiding the cost difference among various inverter systems. Current inverters can use various types of batteries by modifying battery settings.
Assumption 2: The labor costs of building the battery systems are very close and can be neglected.
Assumption 3: Battery system maintenance costs are very close and can be neglected.
Here, current cost information is listed below.
The price of Li-ion batteries per kWh ranges from 500--2000 USD, with an energy density ($D_E$) of roughly about 1/16 kWh/kg.
The price of lead-acid batteries per kWh ranges from 150--500 USD, with an energy density of roughly about 1/35 kWh/kg.
The extra moving cost from the Tataka Saddle (Mt. Jade entrance) to Paiyun Lodge is about 10 USD/kg.

\begin{figure}
\centering
\includegraphics[width=0.8\linewidth,keepaspectratio]{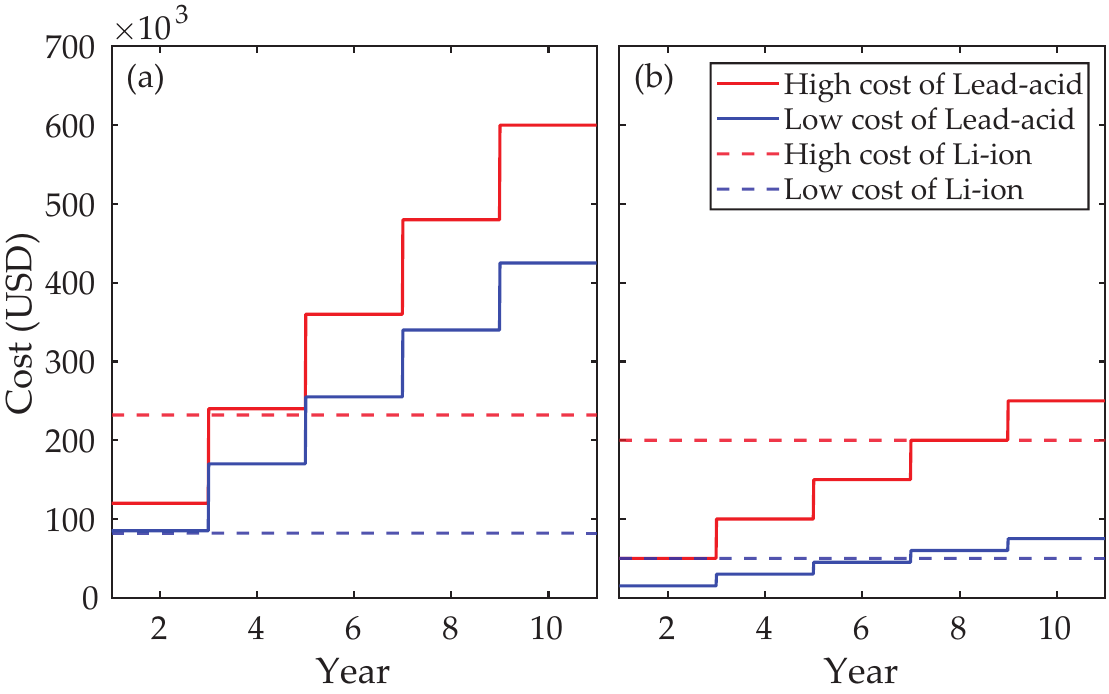}
\caption{
Cost analysis of Li-ion and lead-acid battery systems.
(a) Cost analysis with the extra moving cost included.
The red and blue curves indicate the high and low costs of lead-acid battery systems, respectively.
The red and blue dashed lines indicate the high and low costs of Li-ion battery systems, respectively.
(b) Cost analysis without extra moving costs.
}
\label{fig:CostAnalysis}
\end{figure}

Taking a battery system of total energy $Energy_B = 100$ kWh as an example.
The high and low costs of lead-acid and Li-ion battery systems are shown in Fig.~\ref{fig:CostAnalysis}(a).
For a Li-ion battery system of service life more than 10 years, the high and low costs are the red and blue horizontal dashed lines at 232000 USD and 82000 USD, respectively.
For a lead-acid battery system of service life about 2 years, the high and low costs are the red and blue ladder curves, respectively.
In the first two years, the lead-acid battery is a good choice.
A cross occurs in the second two years (3--4 years), and the Li-ion battery becomes a good choice since the third two years (5--6 years).
It's obvious that for long-term usage of the battery system in high mountains (or a region where extra moving cost is very high), the Li-ion battery system provides a more economical and environmentally friendly choice.
On the other hand, when in a region where there is no extra moving cost, the lead-acid battery will be the cheap choice for a long time and the cross behavior might happen after 8 years as shown in Fig.~\ref{fig:CostAnalysis}(b).
This model reasonably explains that general users will choose lead-acid batteries instead of Li-ion batteries, even if they know that lead-acid batteries are highly polluting.

In the above example, the service life is chosen as 2 years, based on the past battery replacement durations in Paiyun Lodge.
Generally, the service life of lead-acid batteries might range from 2--10 years, which is highly related to the usage conditions, such as C-rate and DOD.
For a car starter battery, the lead-acid batteries are often used at a high C-rate (over $C_R = 3$ h$^{-1}$ for less than 1 second) and very low DOD (less than 1\%), and the service life is 2--5 years~\cite{J.PowerSources42(1993)1P.Ruetschi}.
For a grid-scale energy storage system, the lead-acid batteries are often used under a low C-rate ($C_R < 0.25$ h$^{-1}$) and at a low DOD (less than 50\%), and the service life can even expand to 10--15 years~\cite{J.EnergyStorage15(2018)145G.J.May}.
However, for lead-acid batteries in Paiyun Lodge, the C-rate was low ($C_R < 0.25$ h$^{-1}$) and the DOD was very high (close to 100\%), the service life was largely reduced to roughly 2 years.

Lead-acid batteries were invented in 1859 by French physicist Gaston Plant\'{e}.
It is still one of the most widely utilized battery systems worldwide~\cite{J.PowerSources195(2010)4424P.Kurzweil}.
However, excessive charging will cause lead-acid batteries to emit hydrogen, oxygen, and some toxic gases.
This process is known as ``gassing,'' and toxic gas emissions will result in air pollution.
The sulfuric acid gases can react with the exposed metals, leading to corrosion problems.
Also, lead (Pb) pollution and exposure caused heavy-metal environmental and health problems.
Some lead compounds are extremely toxic.
Long-term exposure to even tiny amounts of lead compounds can result in brain and kidney damage, hearing issues, and learning problems in children~\cite{Environ.Health12(2013)61T.J.Kuijp}.
Poorly managed tailing and gangue disposal processes can carry out water, soil, and transport pollution problems~\cite{Transp.Res.D41(2015)348M.Weiss}.
On the other hand, Li-ion batteries possess no gassing phenomenon during normal working, indicating no on-site toxic air pollution, corrosion, or heavy metal problems.
Hence, Li-ion batteries become eco-friendly replacements for lead-acid batteries.

More than a decade ago, the Li-ion batteries were a pricey production.
The Li-ion battery packs cost 1183 USD/kWh in 2010~\cite{BatteryMarket156USD2019Bloomberg}.
Nine years later, the price had decreased nearly tenfold to 156 USD/kWh in 2019, falling 87\% in real terms~\cite{BatteryMarket156USD2019Bloomberg}.
Cost reductions in 2019 resulted from increasing order size, growth in EV sales, and the continued penetration of high-energy-density cathodes.
The advance in pack designs and falling manufacturing costs will drive prices down.
The prices for Li-ion battery packs across all sectors have reached USD 151/kWh in 2022~\cite{BatteryMarket151USD2022Bloomberg}.
BloombergNEF forecast that Li-ion battery costs will fall under USD 100/kWh in 2024 and hit around USD 60/kWh by 2030~\cite{BatteryMarket151USD2022Bloomberg, BatteryMarket156USD2019Bloomberg}.
The continued low price of Li-ion batteries will make it more likely to replace lead-acid batteries.


\section{Conclusions}

In Li-ion battery analysis, a data processing method has been developed for dealing with large-scale datasets.
Four major daily operation patterns are classified based on the conditions of the solar and backup energy.
The low C-rate distribution ($C_R < 0.15$ h$^{-1}$) made the battery system to avoid from excessive voltage differences among battery cells and heat accumulation on the current collectors, improving system stability and expanding the life of the battery system.
The temperature distributed within the recommended range of operation temperature (15--35$^\circ$C) for more than half of the time, meaning that temperature variation were in normal region and won't affect the system too much.
Also, the temperature difference $\Delta T$ of all time is almost under 5$^\circ$C, indicating that the battery system is in good temperature uniformity.
Overall, the operation temperature and C-rate are close to the minimum of Arrhenius plot, indicating a slower aging rate and longer life of the battery system.
Accumulated capacity (energy) is an important quantity for judging the health of batteries.
Compared to the data of LFP batteries with cycle life about 2000--4000 cycles, the battery system is far away from EOL with an estimated remaining capacity of more than 95\%.
The Li-ion batteries installed in Paiyun Lodge remain healthy and can continue to be used.

The electric power improvement of Paiyun Lodge has been accomplished, making a more durable power system, a more stable battery system, and a more comprehensive EMS.
For the Li-ion battery system, faulty battery cells are replaced, and active balancers for shrinking the voltage differences of battery cells are installed.
For the power system, over-installation of the hybrid solar inverters is applied for system maintenance on high mountains.
Six hybrid solar inverters are installed, when one down, the rest five can still work with load increase by 20\%.
The power of the 6th PV array is reconnected, and bout 16.7\% solar capacity is recovered.
For the EMS, the self-developed cloud EMS has been installed, providing remote monitor of entire energy input and output of the off-grid PV ESS.
On the other hand, the engineering issues on high mountains are given, including (1) alpine worker carrying, (2) demolition and transportation of old and abandoned facilities, (3) material preparation, occupational safety, quality control, safety and health facilities, (4) risks and obstacles.
We believe that after the power improvement project, the re-optimized off-grid PV ESS can provide green electricity for Paiyun Lodge and continue to serve climbers.
Above all, this study gives engineers and researchers a fundamental understanding of a real case of long-term usage of off-grid PV ESSs and engineering on high mountains.

The simple cost model demonstrates that Li-ion batteries (instead of lead-acid batteries) are better choice for long-term application on high mountains.
Owing to the extra moving cost, the cross of the cost curves of two battery systems occurs in the second two years (3--4 years), and the Li-ion battery becomes a good choice since the third two years (5--6 years).
On the other hand, the continued low price of Li-ion batteries in the market will cause the replacement of lead-acid batteries more quickly in the future.



EMS plays an important role in future research.
The old power system possesses no EMS, i.e., no energy data obtained.
It's hard to answer the energy usage problem accurately.
With the installation of EMS in the new power system, we can gather the energy input and output data of the entire power system, evaluating the entire energy input and output accordingly.
PV- and diesel-generated energies, energy consumption, charge and discharge energies of the battery system can be resolved accurately.
The durability of over-installed hybrid inverters and the effect of balancers for long-term usage of the battery system can also be confirmed.
As the annual energy data is obtained, we can realize whether the capacities of PV and ESS are sufficiently large or not.
If the PV and ESS capacities are not sufficient, we will enlarge the capacities accordingly.
Our goal is to largely reduce GHG emissions by using renewable energies.

\vspace{6pt}

\funding{This work was supported in part by Yushan National Park Headquarters, Taiwan under the contract number: 111-d08.
This work was supported in part by Super Double Power Technology Co., Ltd., Taiwan, under the project ``Development of Cloud-native Energy Management Systems for Medium-scale Energy Storage Systems (\href{https://osf.io/7fr9z/}{https://osf.io/7fr9z/})'' (Grant number: SDP-RD-PROJ-001-2020).}

\dataavailability{The data are not publicly available due to privacy constraints.}




\acknowledgments{The author (H. C. Chung) would like to thank the contributors to this article for their valuable discussions and recommendations, Jung-Feng Jack Lin, Hsiao-Wen Yang, Yen-Kai Lo, and An-De Andrew Chung.
The author (H. C. Chung) thanks Pei-Ju Chien for English discussions and corrections as well as Ming-Hui Chung, Su-Ming Chen, Lien-Kuei Chien, and Mi-Lee Kao for financial support.}

\conflictsofinterest{The authors declare no conflicts of interest.}

%
%
%

\abbreviations{Symbols}{
The following symbols are used in this manuscript:\\

\noindent
\begin{tabular}{@{}ll}
$C_{acc}$           & Accumulated capacity\\
$Cap_N$             & Nominal ampere-hour capacity\\
$C_{day}$           & Daily accumulated capacity\\
$C_{day}^{chg}$     & Daily accumulated charge capacity\\
$C_{day}^{dis}$     & Daily accumulated discharge capacity\\
$C_{year}^{chg}$    & Annual accumulated charge capacity\\
$C_{year}^{dis}$    & Annual accumulated discharge capacity\\
$C_{nom}$           & Nominal capacity\\
$Cost$              & Total cost of the battery system\\
$Cost_B$            & Battery cost\\
$Cost_M$            & Moving cost due to high mountains\\
$C_R$               & C-rate\\
$CR_{year}^{loss}$  & Annual capacity loss rates\\
$D_E$               & Energy density of the battery system\\
$\Delta C_{day}$    & Difference between the daily accumulated charge and discharge capacities\\
$\Delta E_{day}$    & Difference between the daily accumulated charge and discharge energies\\
$\Delta T$          & Difference between maximum and minimum battery cell temperatures\\
$\Delta t_n$        & $n$th interval\\
$\Delta V_{avg}$    & Average voltage difference\\
$E_{acc}$           & Accumulated energy\\
$E_{day}$           & Daily accumulated energy\\
$E_{day}^{chg}$     & Daily accumulated charge energy\\
$E_{day}^{dis}$     & Daily accumulated discharge energy\\
$E_{year}^{chg}$    & Annual accumulated charge energy\\
$E_{year}^{dis}$    & Annual accumulated discharge energy\\
$Energy_B$          & Total energy of the battery system\\
$ER_{year}^{loss}$  & Annual energy loss rates\\
$I$                 & Current\\
$I(t_n)$            & Current value of the $t_n$ point\\
$Price_B$           & Battery price per kWh\\
$Price_M$           & Moving price per kg\\
$V$                 & Voltage\\
$V_{cell}$          & Voltage of the battery cell\\
$V_n$               & Voltage of the $n$th battery cell\\
$V(t_n)$            & Voltage value of the $t_n$ point\\
$V_{tot}$           & Total voltage\\
$T$                 & Temperature\\
$T_{max}$           & Maximum temperature of the battery cells in pack\\
$T_{min}$           & Minimum temperature of the battery cells in pack\\
$T_n$               & Temperature detected from the $n$th sensor\\
$t_n$               & A point in the partition of $[a, b]$\\
$^\circ$C           & Degree Celsius\\
\end{tabular}
}

\abbreviations{Abbreviations}{
The following abbreviations are used in this manuscript:\\

\noindent
\begin{tabular}{@{}ll}
1P2W           & One-phase-two-wire\\
3C             & Computing, communication, and consumer\\
3D             & Three-dimensional\\
AC             & Alternative current\\
AMS            & Acute mountain sickness\\
BCE            & Before the Common Era\\
BESS           & Battery energy storage system\\
BMS            & Battery management system\\
CC             & Constant current\\
CC-CV          & Constant current-constant voltage\\
CSV            & Comma-separated values\\
DC             & Direct current\\
DDoS           & Distributed denial of service\\
DOD            & Depth of discharge\\
ECM            & Equivalent circuit model\\
EMS            & Energy management system\\
EOL            & End of life\\
ESS            & Energy storage systems\\
EV             & Electric vehicle\\
FCAS           & Frequency control ancillary services\\
GHG            & Greenhouse gas\\
HACE           & High-altitude cerebral edema\\
HAPE           & High-altitude pulmonary edema\\
IEA            & International Energy Agency\\
LED            & Light-emitting diode\\
LFP battery    & Lithium iron phosphate battery\\
Li-ion battery & Lithium-ion battery\\
LLSS           & Lake Louise scoring system\\
LNT            & Leave No Trace\\
MATLAB         & Matrix laboratory\\
MPPT           & Maximum power point tracker\\
PV             & Photovoltaic\\
SEI            & Solid electrolyte interface\\
USD            & United States dollar
\end{tabular}
}

%
%

\begin{adjustwidth}{-\extralength}{0cm}

\reftitle{References}

\PublishersNote{}
\end{adjustwidth}
\end{document}